\documentclass[]{emulateapj}

\usepackage{natbib}
\usepackage{amsmath}
\usepackage{epsfig}
\usepackage{graphicx}
\usepackage{color}
\newcommand{\rxj}{RX J$\,1347.5{-}1145$}

\shorttitle{High Spectral Resolution Measurement of the SZ Null with
  Z-Spec} \shortauthors{ZEMCOV ET AL.}

\begin{document}

\slugcomment{Received 2011 September 2; accepted 2012 February 3; published 2012 March 29.}

\title{HIGH SPECTRAL RESOLUTION MEASUREMENT OF THE SUNYAEV--ZEL'DOVICH EFFECT NULL WITH Z-Spec}

\author{M. Zemcov\altaffilmark{1,2}, 
  J. Aguirre\altaffilmark{3},
  J. Bock\altaffilmark{2,1}, 
  C.~M. Bradford\altaffilmark{2,1},
  N. Czakon\altaffilmark{1},
  J. Glenn\altaffilmark{4},
  S. R. Golwala\altaffilmark{1},
  R. Lupu\altaffilmark{3}, 
  P. Maloney\altaffilmark{4},
  P. Mauskopf\altaffilmark{5},
  E. Million\altaffilmark{6},
  E.~J. Murphy\altaffilmark{7},
  B. Naylor\altaffilmark{1,2},
  H. Nguyen\altaffilmark{2,1}, 
  M. Rosenman\altaffilmark{3},  
  J. Sayers\altaffilmark{1},
  K.~S. Scott\altaffilmark{8}, and
  J. Zmuidzinas\altaffilmark{1,2}
}

\altaffiltext{1}{Department of Physics, Mathematics and Astronomy, California Institute of Technology, Pasadena, CA 91125, USA; zemcov@caltech.edu}
\altaffiltext{2}{Jet Propulsion Laboratory (JPL), National Aeronautics and Space Administration (NASA), Pasadena, CA 91109, USA}
\altaffiltext{3}{Department of Physics and Astronomy, University of Pennsylvania, Philadelphia, PA 19104, USA}
\altaffiltext{4}{Center for Astrophysics and Space Astronomy, University of Colorado, CASA 389-UCB, Boulder, CO 80303, USA}
\altaffiltext{5}{School of Physics and Astronomy, Cardiff University,
  Cardiff, CF24 3YB, UK}
\altaffiltext{6}{Department of Physics and Astronomy, University of Alabama,, 206 Gallalee Hall Box 870324, Tuscaloosa AL 35487, USA}
\altaffiltext{7}{Observatories of the Carnegie Institution for Science, 813
Santa Barbara Street, Pasadena, CA 91101, USA}
\altaffiltext{8}{North American ALMA Science Center, National Radio Astronomy
Observatory, Charlottesville, VA 22901}


\begin{abstract}
The Sunyaev--Zel'dovich (SZ) effect spectrum crosses through a null
where $\Delta T_{\mathrm{CMB}}=0$ near $\nu_{0}=217 \,$GHz.  In a
cluster of galaxies, $\nu_{0}$ can be shifted from the canonical
thermal SZ effect value by corrections to the SZ effect scattering due
to the properties of the inter-cluster medium.  We have measured the
SZ effect in the hot galaxy cluster \rxj\ with Z-Spec, an $R \sim 300$
grating spectrometer sensitive between 185 and $305 \,$GHz.  These
data comprise a high spectral resolution measurement around the null
of the SZ effect and clearly exhibit the transition from negative to
positive $\Delta T_{\mathrm{CMB}}$ over the Z-Spec band.  The SZ null
position is measured to be $\nu_{0} = 225.8 \pm 2.5 \mathrm{(stat.)}
\pm 1.2 \mathrm{(sys.)} \,$GHz, which differs from the canonical null
frequency by $3.0 \sigma$ and is evidence for modifications to the
canonical thermal SZ effect shape.  Assuming the measured shift in
$\nu_{0}$ is due only to relativistic corrections to the SZ spectrum,
we place the limit $k T_{\mathrm{e}} = 17.1 \pm 5.3 \,$keV from the
zero-point measurement alone.  By simulating the response of the
instrument to the sky, we are able to generate likelihood functions in
$\{y_{0}, T_{\mathrm{e}}, v_{\mathrm{pec}}\}$ space.  For
$v_{\mathrm{pec}} = 0 \,$km s$^{-1}$, we measure the best fitting SZ
model to be $y_{0} = 4.6^{+0.6}_{-0.9} \times 10^{-4},
T_{\mathrm{e},0} = 15.2^{+12}_{-7.4} \,$keV.  When $v_{\mathrm{pec}}$
is allowed to vary, a most probable value of $v_{\mathrm{pec}} =
{+}450 \pm 810 \,$km s$^{-1}$ is found.
\vspace{0.4cm}
\end{abstract}

\keywords{cosmic background radiation -- galaxies: clusters:
  individual (\rxj) -- galaxies: clusters: intracluster medium --
  submillimeter: galaxies}


\section{Introduction}
\label{S:intro}

\setcounter{footnote}{0}

\begin{figure}[ht]
\centering
\epsfig{file=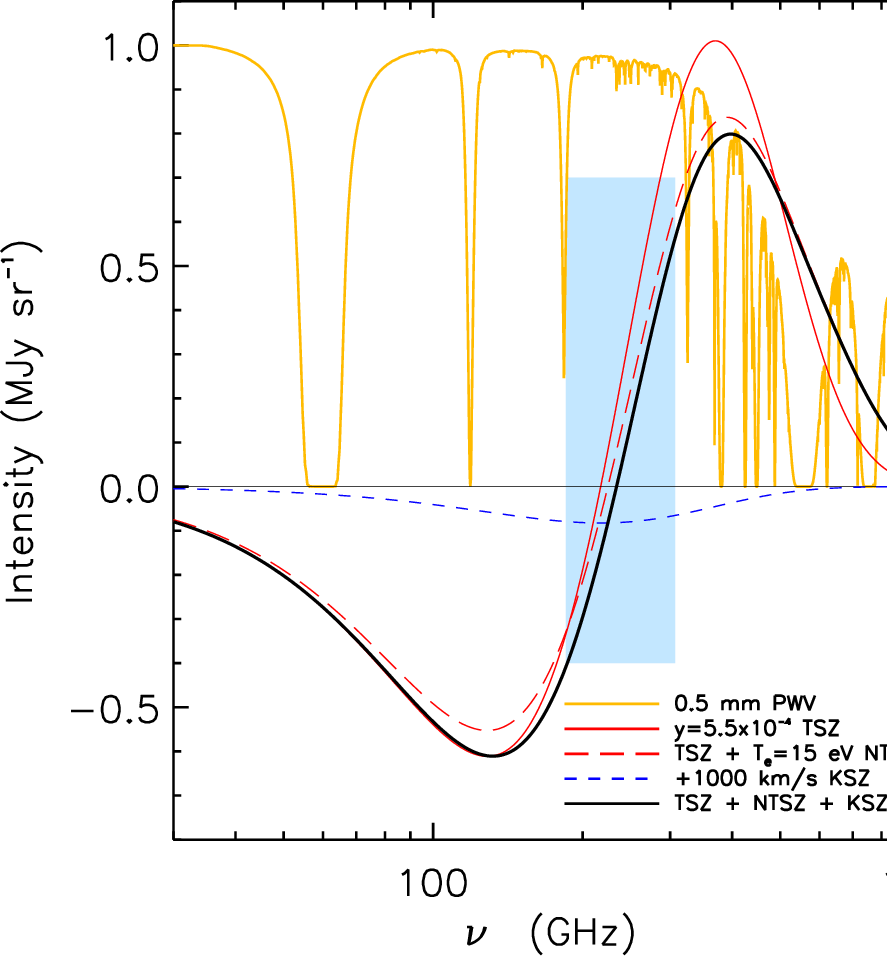,width=0.4\textwidth}
\caption{SZ effect from 30 to $1000 \,$GHz.  The pure thermal SZ
  effect with a canonical null at $\nu_{0} = 217 \,$GHz is shown as a
  solid red line, the thermal effect including non-thermal corrections
  as a dashed red line, the kinetic effect for a cluster velocity of
  $v_{\mathrm{pec}} = + 1000 \,$km s$^{-1}$ as a blue (short) dashed line,
  and the sum of all three effects --- which is the spectrum which would
  be observed through the cluster --- as the solid black line.  Also
  shown are the atmospheric transmittance spectrum for a precipitable
  water vapor column of $0.5 \,$mm at Mauna Kea in yellow (P.~Ade 2007,
  private communication), and the effective bandwidth of Z-Spec on
  the CSO in the solid blue region.  Z-Spec's spectral range is well
  matched to span the null in the SZ effect.}
\label{fig:defaultSZspectrum}
\end{figure}

The Sunyaev--Zel'dovich (SZ) effect is a distortion in the blackbody
spectrum of the cosmic microwave background (CMB) radiation caused by
inverse Compton scattering of CMB photons from hot electrons in a
plasma such as that found in intracluster media (ICM; \citealt{SZ1972}).
Given the dimensionless frequency $x = k T_{\mathrm{CMB}} / h \nu$,
the ICM gas temperature $T_{\mathrm{e}}$, and a cluster's velocity
along the line of sight with respect to the CMB $v_{\mathrm{pec}}$,
the intensity shift from the CMB background due to the SZ effect can
be expressed as
\begin{equation}
\Delta I = y \left| \frac{d B_{\nu}}{d T} \right| \left[ (f(x) (1 +
  \delta(x,T_{\mathrm{e}})) +
  k_{\mathrm{SZ}}(x,y,T_{\mathrm{e}},v_{\mathrm{pec}}) \right]
\label{eq:1}
\end{equation}
where $y$ is the Compton parameter proportional to the sight line
integrated ICM pressure, $f(x) = x (e^{x} + 1) / (e^{x} - 1) - 4$,
$d B_{\nu} / d T$ is the derivative of the Planck function at the
temperature of the CMB, and $\delta$ and $k_{\mathrm{SZ}}$ are
corrections due to the relativistic and kinetic SZ effects,
respectively (\citealt{Birkinshaw1999}; \citealt{Carlstrom2002}).
Figure \ref{fig:defaultSZspectrum} shows the various contributions to
the SZ effect for a typical massive galaxy cluster.

At most frequencies, the largest contribution to the sum in Equation
(\ref{eq:1}) is from the thermal SZ (tSZ) effect, given by $\Delta
I_{\mathrm{thermal}} = y f(x) \left| d B_{\nu} / d T \right|$.  A
characteristic property of the tSZ effect is that, because on average
CMB photons gain energy from the hot electrons, a decrement in
temperature compared to the CMB arises at low frequencies, and an
increment arises at high frequencies.  The crossover between the
increment and decrement occurs at $\nu_{0} = 217 \,$GHz and is the
``null'' of the SZ effect.

The relativistic SZ (rSZ) effect $\delta(x,T_{\mathrm{e}})$ in
Equation (\ref{eq:1}) arises because a full description of the
scattering properties of the ICM gas requires corrections due to the
relativistic velocities of very hot electrons in the ICM.  These
corrections are typically small in the Rayleigh--Jeans part of the
spectrum, but become large near the SZ effect peaks and null: for
example, at the maximum of the SZ increment, this correction changes
the SZ flux by $18 \,$\% for a $k T_{\mathrm{e}} = 15 \,$keV cluster
(\citealt{Wright1979}; \citealt{Challinor1998}; \citealt{Itoh1998};
\citealt{Nozawa2000}).  The shift in the position of the null for such
a cluster is $\Delta \nu / \nu \approx 5\,$\%, which is substantial;
more importantly, as the non-thermal corrections depend on $x$, their
effect on the spectrum also changes with observing frequency.  This
means that a sufficiently accurate measurement of the SZ effect
spectrum at different frequencies can allow a measurement of the
amplitude of the relativistic corrections to the tSZ effect
(\citealt{Hansen2002}, \citealt{Zemcov2010}).

In contrast, the kinetic SZ (kSZ) effect $k(v_{\mathrm{pec}})$ has the
same spectral shape as the CMB but scales according to the
line-of-sight peculiar velocity of the cluster.  Importantly, the kSZ
spectrum does not have an intensity null, but rather is entirely
positive or negative depending on the velocity of the cluster.  For
reasonable cluster velocities given structure formation models ---
usually $\approx 1000 \,$km s$^{-1}$ --- the ratio of the $217 \,$GHz
kinetic SZ effect brightness and the peak increment brightness due to
the combined thermal and non-thermal effects is $\lesssim 10 \,$\%.
Though faint, such levels can be reached with the current generation
of instruments \citep{Benson2003}.

As tSZ effect probes only the pressure of the ICM, the rSZ corrections
and kinetic SZ effect allow more detailed studies of the ICM.
However, due to degeneracies between the tSZ spectrum and both the rSZ
and kSZ spectra, it is often difficult to disentangle all three
components, even with multi-band data. This is especially true in the
presence of cluster substructure.  Though difficult to measure, the
rSZ effect can be used to study the temperature of the ICM, especially
at the high temperatures that are difficult to constrain with current
X-ray facilities, and at high redshifts where X-rays are attenuated by
cosmological dimming.  For example, \citet{Prokhorov2011} show that a
comparison of SZ images at different frequencies can allow
measurements of the morphology of the temperatures in ICMs through the
changing emission from the rSZ corrections. Although similarly
difficult to measure, the kSZ effect can be used to study cosmology by
constraining the peculiar velocites of clusters, and to study cluster
astrophysics by constraining the velocities of bulk flows within the
cluster itself (see \citep{Birkinshaw1999} for a comprehensive
review). In addition, the kSZ and rSZ signals can be used together to
constrain complex temperature and velocity substructures due to
merging activity in clusters (\citealt{Diego2003}; \citealt{Koch2004};
\citealt{Prokhorov2010}).  In either case spectral information,
preferably spanning the null in the SZ effect, is key.

Z-Spec is a multi-channel spectrometer working in the $220 \,$GHz
atmospheric window with 160 independent spectral channels running from
185 to $305 \,$GHz \citep{Bradford2004}.  Z-Spec's spectral range
almost ideally spans the SZ effect null and so provides a unique
opportunity to measure the transitional frequencies where the tSZ
effect passes from a decrement to an increment in surface brightness
with high spectral resolution.  In principle, such measurements can
tightly constrain the position of the SZ null as well as the presence
of any foreground or background emission via measurements of line
emission.  In this paper, we present millimeter (mm)/submillimeter
(sub-mm) Z-Spec measurements spanning the SZ null in the massive,
SZ-bright galaxy cluster \rxj.  Additional Bolocam \citep{haig04} data
are used to build a model of the spatial shape of the SZ emission in
the cluster, and to help constrain the kSZ corrections in it.  The
paper is structured as follows: a description of the observations
performed is presented in Section \ref{S:observations}, and the data
analysis procedure is described in Section \ref{S:analysis}.  Sections
\ref{S:results} and \ref{S:discussion} present the results and the
simulations required to interpret them, and finally a discussion of
the implications of this work, respectively.

\section{Observations}
\label{S:observations}

Both Z-Spec and Bolocam observations are used in this work and are
detailed below.  Though this paper primarily addresses measurement of
the spectral shape of the SZ effect with Z-Spec, a spatial model of
the cluster is necessary to interpret the spectral measurement, for
which we use the Bolocam maps.  Bolocam also adds a lower frequency
datum with small uncertainties to the SZ spectrum which is used in
Section \ref{sS:vpec}.

\subsection{Z-Spec Observations}
\label{sS:ZSobs}

Descriptions of the instrumental design of Z-Spec and the tests used
to characterize its performance are described in, e.g.,
\citet{Bradford2004}, \citet{Naylor2008}, and \citet{Bradford2009};
here we review the details salient to this work.  Z-Spec is a grating
spectrometer which disperses the input light across a linear array of
160 bolometers.  The curved grating operates in a parallel plate
waveguide fed by a single-moded feed horn, which is coupled to the
telescope via specialized relay optics.  As with all grating
spectrometers, the spectral range of the instrument is set by the
geometry of the input feed, grating, and detector elements; for the
Z-Spec configuration discussed in this work, the spectral range is
measured as $185 \,$GHz $\leq \nu \leq 307 \,$GHz.  The resolving
power of the grating varies across the linear array, providing
single-pixel bandwidths of $\Delta \nu = 500 \,$MHz at the
low-frequency end of the range to $\Delta \nu = 1200 \,$MHz at the
high-frequency end with a mean value of $\Delta \nu = 750 \,$MHz.
Measurements of the spectral bandpass of each of the detectors on the
ground and measurements of line positions of bright astronomical
sources yield an uncertainty on the central bandpass of the detectors
less than $200 \,$MHz.  In order to achieve photon background limited
performance, the detectors, grating and input feed horn are housed in
a cryostat and the entire assembly is cooled with an adiabatic
demagnetization refrigerator to between 60 and $85\,$mK.  The
detectors, which are designed and fabricated at JPL, have noise
equivalent powers of $4 \times 10^{-18} \,$W Hz$^{-1/2}$ which make
them the most sensitive, lowest-background detectors used for
astrophysics to date.

Z-Spec has been operating as a PI instrument at the Caltech
Sub-millimeter Observatory (CSO) on Mauna Kea, Hawaii since 2007.
Z-Spec employs a chop and nod technique to remove the in-band
atmospheric emission, which is the largest source of time-correlated
noise in the system.  Since it is not possible to modulate the
spectral response of the instrument, the chopping secondary mirror of
the telescope is used to provide on-sky modulation.  In general, there
is non-negligible SZ emission beyond the virial radius of massive
galaxy clusters, typically corresponding to 1--2~Mpc or a few
arcminutes at moderate $z$.  This means that chopping and nodding
using throws designed for point source observations will typically not
be large enough to put the telescope beam off the cluster.  The
presence of this signal in the off-beams in turn attenuates the
observed signal since non-zero astronomical flux is differenced from
the main observation beam as part of the signal modulation.  In order
to mitigate this effect, a target which is bright, compact, and does
not have sub-mm bright galaxies in the measurement or chopping beams
is desirable.  \rxj\ at $z=0.451$, one of the brightest known X-ray
and SZ clusters, fulfills these criteria.  The SZ emission in this
cluster has been mapped at several frequencies in the past including
in the Rayleigh--Jeans regime (e.g., \citealt{Reese2002};
\citealt{LaRoque2006}), at the peak of the decrement
(e.g., \citealt{Komatsu2001}; \citealt{Pointecouteau2001}), and near
the peak of the increment (e.g., \citealt{Komatsu1999};
\citealt{Zemcov2007}).  The bulk SZ emission in this cluster is
unusually centrally peaked and bright with a measured $y_{0} \sim 5
\times 10^{-4}$ and $k T_{\mathrm{e}} > 15.0 \,$keV
\citep{Bradac2008}; this means that the expected SZ emission will have
exceptionally high contrast between the center and chop positions.  In
addition, two especially SZ-bright regions associated with shocked gas
falling into the cluster lie $\sim 10 \,$arcsec away from its center
(\citealt{Komatsu2001}; \citealt{Mason2010}).  Compared to other
clusters the foreground and background contamination in \rxj\ is
slight; the maps of \citet{Zemcov2007} show that there is only a
single bright (presumably lensed) sub-mm source originally reported by
\citet{Kitayama2004} which could contaminate the SZ effect
measurement, but this is easily avoided using a suitable chop throw on
the sky.  Additionally, works like that of \citet{Pointecouteau2001}
and \citet{Mason2010} show that while the central cluster galaxy's
ratio flux is large, it is expected to contribute only $1{-}2 \,$mJy
over the Z-Spec band; this source also can be avoided by a suitable
observation strategy.

\rxj\ was observed with Z-Spec on the CSO 2012 March 7 through 12,
with a total on-target time of 12.1 hr ($=43.6 \,$ks).  Z-Spec was
pointed toward the SZ-bright shock centered at $13^{\mathrm{h}}
47^{\mathrm{m}} 31^{\mathrm{s}}.4, -11^{\circ} 45' 25''$ (J2000).  The
weather as measured by the atmospheric opacity $\tau_{225
  \mathrm{GHz}}$ was excellent; $\tau_{225 \mathrm{GHz}}$ had a mean
of $0.034$ with a standard deviation of $0.014$ over the six nights of
observation.  Pointing and spectral flat-field observations bracketed
the science observations for each night; these were taken using the
radio bright sources 3C279 and 3C345 for whose spectral
characteristics are well understood.  Calibration observations were
taken using Mars, the standard Z-Spec calibration source.  As
discussed in \citet{Bradford2009}, the Z-Spec absolute calibration is
determined by a combination of planetary observations and the
empirically determined relationship between each bolometer's operating
voltage and responsivity.  Based on the consistency of these
measurements spanning several years of observations, the channel to
channel calibration uncertainties are less than 10 \% except at the
very lowest frequencies where the $186 \,$GHz water line is a strong
contaminant.  Each observation was taken using a chop and nod
technique with a throw of $90 \,$arcsec and a chop frequency of $f_{0}
= 1.62 \,$Hz; nods occurred with a frequency of $n_{0} = 10.44 \,$mHz.

\subsection{Bolocam Observations}
\label{sS:BCobs}

Bolocam is a large format camera at the CSO with 144 detectors, an 8
arcmin diameter field of view (FOV), an observing band centered at
2.1~mm, and a point-spread function with a 58~arcsec full-width at
half-maximum (FWHM; \citealt{haig04}).  As part of an ongoing program to
image the SZ effect in an X-ray-selected sample of massive galaxy
clusters, we observed \rxj~with Bolocam from the CSO in 2008 July and
2009 May.  The total on-source integration time is $59 \,$ks,
approximately evenly split between the two observing periods.  The
central rms of the beam-smoothed Bolocam data is
16~$\mu$K$_{\textrm{CMB}}$, and the peak decrement of \rxj~is $\simeq
850$~$\mu$K$_{\textrm{CMB}}$.

\begin{figure}[ht]
\centering
\epsfig{file=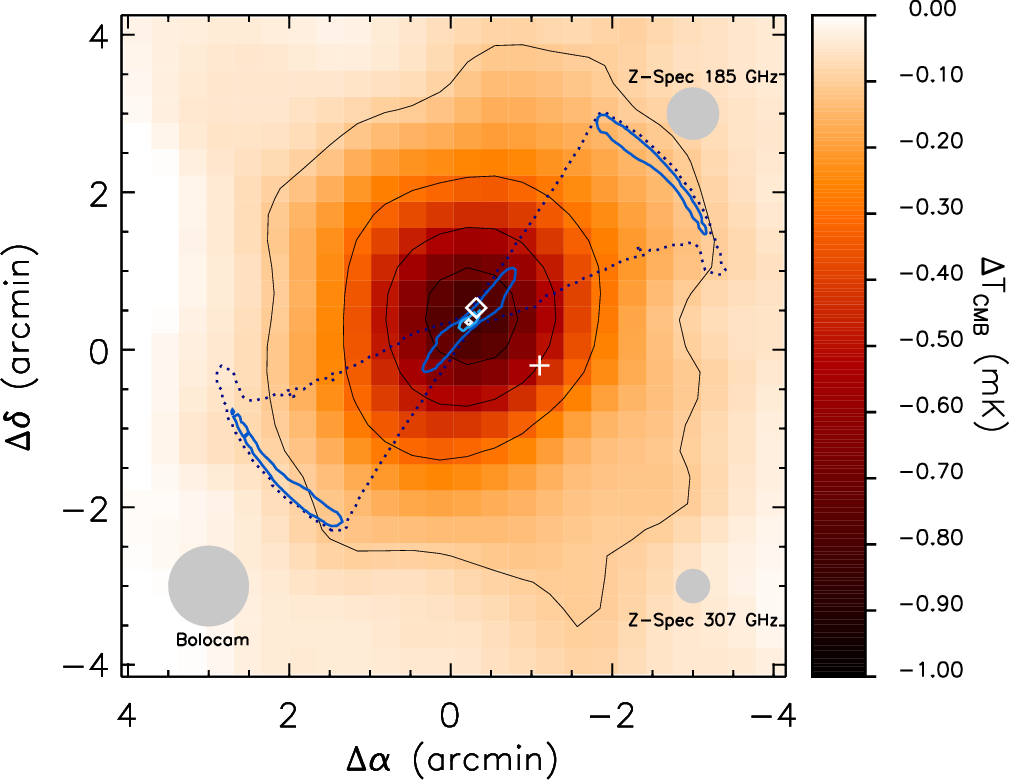,width=0.4\textwidth}
\caption{Bolocam 140~GHz image of \rxj\ with the Z-Spec observing
  scheme superimposed. The Bolocam image has been beam-smoothed to an
  effective resolution of 82~arcsec, but is otherwise unfiltered
  (i.e., the effects of noise filtering have been deconvolved). The
  central rms per beam in the image is 16 $\mu$K$_{\textrm{CMB}}$.  The
  blue contours show how the Z-Spec observations were performed; the
  central point is the ``on-source'' beam while the extended lobes
  show the effective integration time on the two chop/nod positions.
  Integrating the chop lobes would yield the same integration time as
  was spent on the cluster.  This figure highlights the traditional
  difficulty with SZ observations using a chop and nod scheme, which
  is that it is difficult to chop the beam of the instrument far
  enough to fall on a region with small SZ amplitude.  Given Z-Spec's
  90 arcsec chop throw we naively expect a reduction in the measured
  SZ amplitude by a factor of $\sim 0.5$.  In addition, we have
  indicated the positions of the central AGN in this cluster with a
  diamond symbol, and the position of the known sub-mm bright source
  with a cross.  Finally, the angular size of the FWHM of the beams of
  both Bolocam and the extreme high and low frequencies of Z-Spec
  are shown as gray circles at the edges of the image. }
\label{fig:RXJ1347_chop}
\end{figure}

The Bolocam observations were made by scanning the CSO in a Lissajous
pattern, with an average scan speed of $4 \,$arcmin s$^{-1}$.  All of
the data were reduced following the procedures described in detail in
\citet{sayers11}.  In particular, atmospheric noise was subtracted by
a combination of an FOV-average template and a time stream high pass
filter.  A pointing model, accurate to 5~arcsec, was created from
frequent observations of the nearby bright object $1334{-}127$
following the methods described in \citet{sayers09}.  The flux
calibration, in nV/$\mu$K$_{\textrm{CMB}}$, was determined according
to the procedure described in \citet{laurent05}, based on observations
of Neptune, Uranus, and G34.3 using the brightness values determined
in \citet{sayers11prep}.  We estimate the flux calibration to be
accurate to $\simeq 5$\%~\citep{sayers11, sayers11prep}.  For the
analysis presented in this paper, an elliptical isothermal $\beta$
model was fit to the Bolocam data following the procedures described
in detail in \citet{sayers11}; the best-fitting parameters of this
model are given in Table \ref{tab:isothermalbeta}.  The Bolocam map of
\rxj\ with the Z-Spec chopping pattern overlaid is shown in Figure
\ref{fig:RXJ1347_chop}.

\newpage

\section{Data Analysis}
\label{S:analysis}

The standard Z-Spec data analysis pipeline is used to account for
instrumental effects, perform the calibration from voltage to
astronomical flux, and rectify the chop modulation.  In addition,
customized data cuts and simulations are required to interpret these
data.  We describe this data reduction and simulation work in detail
in this section.

\subsection{Data Reduction}
\label{sS:reduction}

The Z-Spec data reduction pipeline is designed to extract astronomical
signals from the measured data time lines which include a variety of
instrumental effects as well as statistical noise; a detailed
description of this pipeline can be found in \citet{Naylor2008}.  The
basic building blocks of the data analysis are individual data files
consisting of 20 nods or $33$ minutes of data; these contain the
merged raw time-ordered data (TOD), telescope pointing, chop and nod
positions, and general housekeeping data from the telescope.  

The first step in the reduction pipeline is to deglitch the TOD which
is performed by applying a simple $\sigma$-clip based on statistics
calculated from $1000$ sample long windows; all points more than $3
\sigma$ from the mean of the time line for each bolometer are masked.
These deglitched data are then low pass filtered using a Kaiser window
with a pole at $4.5 \,$Hz and are down sampled by a factor of three.
Since the average of Z-Spec TOD does not reflect the absolute value of
the astronomical sky, each raw time line is then mean subtracted to
yield zero mean over the full chop/nod cycle in each TOD set.

The data files contain the raw secondary mirror encoder which monitors
the position of the chop; this occurs at a predictable rate with a
uniform though complex waveform.  To simplify the analysis, the
chopper data are filtered and down sampled in the same way as for the
TOD and an artificial pure tone modulation function phase locked with
the chopper signal is generated.  An additional complexity is the
possible presence of phase shifts between the chopper encoder and the
bolometer readout.  Calibration observations of bright sources are
used to measure and correct the pure tone rectification signal for
this phase delay, which are found to be stable over a night of
observing.  Given the phase delay, the per nod TODs are demodulated
using the appropriate pure tone model to yield the signal accounting
for the chop pattern, and the power in single half nod (i.e., over the
``A'' or ``B'' component of the standard ABBA-style difference) is found.
These demodulated averages are then subtracted from one another using
the standard ABBA formulation \citep{Zemcov2003} to yield the
measurement for each nod cycle.  These averages, which each comprise
the average of a $100 \,$s long measurement of the spectrum of the sky
for each bolometer, are then passed to the next stage of processing.

In order to determine the uncertainty associated with each nod
measurement, the white noise level of the time streams is measured.
The power spectral density (PSD) of the raw per nod time streams is
generated, and a window function defined by $1.4 \,$Hz $< f < 2.5
\,$Hz is generated.  This window is designed to sit above the $1/f$
knee in the data and below the second harmonic of $f_{0}$.  The noise
level in the PSD over this window excluding the points about the first
harmonic of $f_{0}$ is calculated and assigned as the uncertainty in
the data for that nod.  This is repeated for all nods in the data set.

\begin{figure*}[ht]
\centering
\epsfig{file=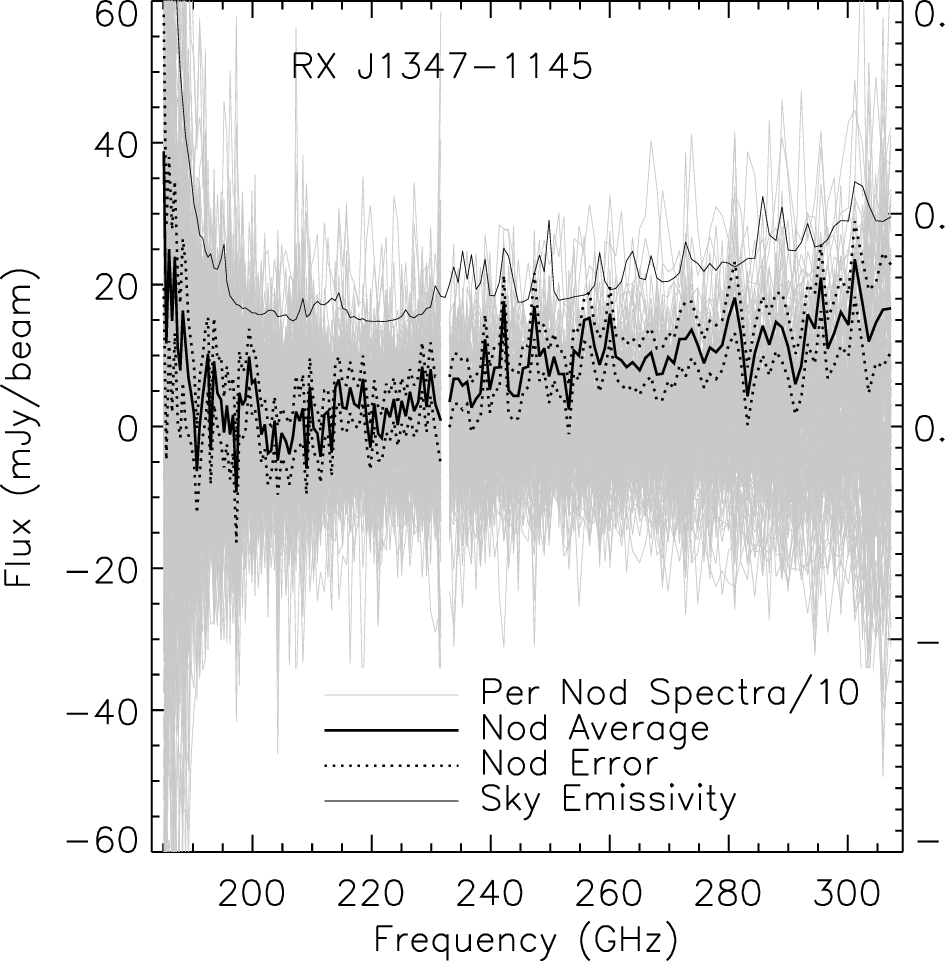,width=0.4\textwidth}
\hspace{1.2cm}
\epsfig{file=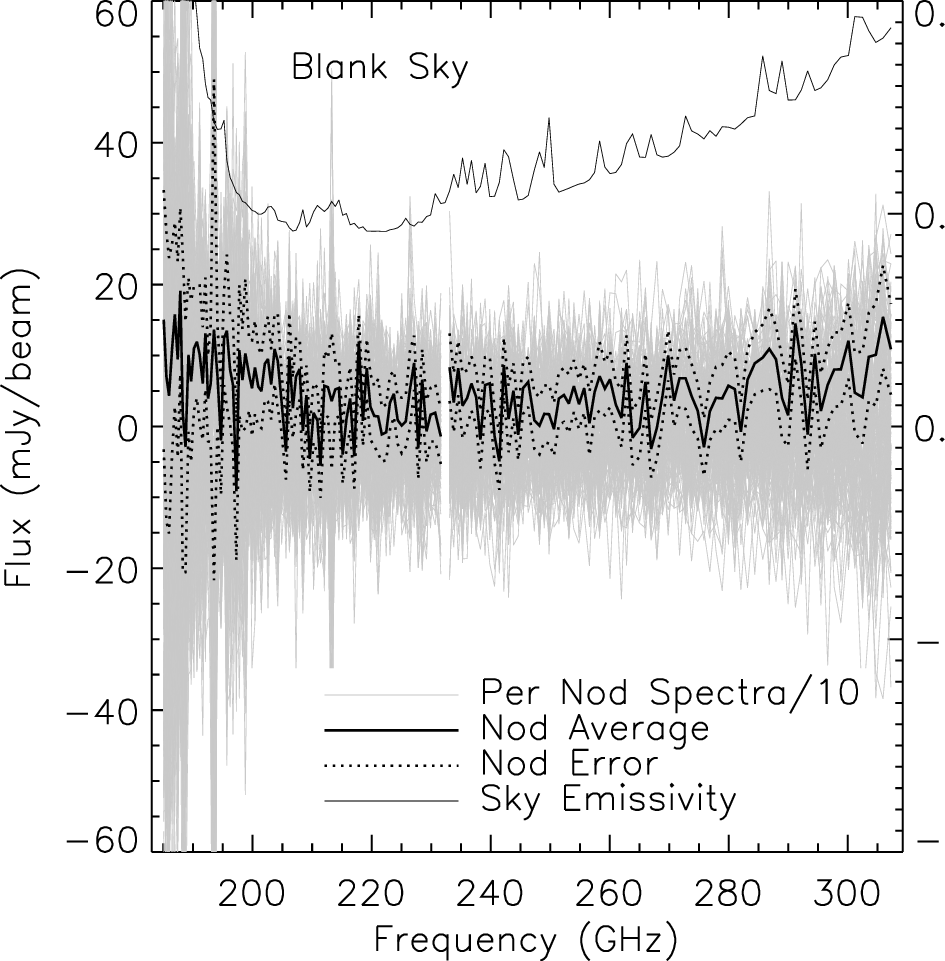,width=0.4\textwidth}
\caption{Raw nod spectra, averages and sky emissivity for the
  \rxj\ (left panel), and blank sky (right panel) fields.  Each plot
  shows the set of raw nod spectra as measured by Z-Spec (gray lines)
  and the per bolometer variance-weighted mean (black solid lines), and
  error on the mean (black dashed lines).  The raw nod spectra have
  been scaled by a factor of 10 to allow them to be plotted on the
  same axis as the mean.  Plotted on the same panel scaled to the
  right hand axis is the mean sky emissivity $\epsilon$ for both
  targets.  The nod averages are strongly correlated with $\epsilon$
  for each bolometer, which is evidence that the atmospheric
  emissivity is varying on short timescales and summing to affect the
  behavior over a nod cycle.}
\label{fig:allnods}
\end{figure*}

To produce the final sky spectrum, all of the individual nods must be
combined.  This is performed using a noise-weighted average, where the
noise weights are the variance derived from the PSD.  Figure
\ref{fig:allnods} shows the spectra measured in each individual nod
for the \rxj\ observations and the noise-weighted co-addition of the
nods.  Bolometer 81, monitoring the $232 \,$GHz channel, is known to
have no optical response and is removed from further analysis.  Also
plotted in Figure \ref{fig:allnods} is the measured emissivity of the
atmosphere in the Z-Spec band during these observations; the co-added
nod spectrum and the atmospheric emissivity are strongly correlated.
Investigation of the individual nods shows that in many cases,
particularly during periods of large noise, the nods have the shape of
the atmospheric emission.  This is due to gradients in the overall
amplitude of the sky brightness temperature which cause the
atmospheric signal to decorrelate on timescales faster than the chop
($f \gtrsim 1 \,$ Hz).  Though generally below the noise floor per
chop cycle, these gradients add constructively over many chop cycles
and can cause significant nod cancellation failure.  Examination of
the individual nod-differenced spectra shows that only a fraction of
the nods show structure correlated with the atmospheric emission,
which suggests that those spectra showing such structure can be found
and masked from the co-addition algorithmically.

Z-Spec data are available which allow blind construction of a data
masking algorithm.  Z-Spec observed a random dark spot on the sky
centered at $17^{\mathrm{h}} 19^{\mathrm{m}} 10^{\mathrm{s}}.3,
{+}58^{\circ} 55' 54''.2$ on 2011 May 11 and 12 for a total of $6.6
\,$hr ($=23.6 \,$ks) using the same azimuth chop and nod scheme as were
used in the \rxj\ observations.  Figure \ref{fig:allnods} also shows
the individual nod spectra, noise-weighted average spectrum, and
atmospheric emissivity for the blank sky observations.  Here, the
correlation with the atmospheric emissivity is slightly less
pronounced than in the \rxj\ observations, though still present.
\citet{Sayers2010} present a detailed analysis of the fluctuation
properties of the atmosphere from the CSO in a band centered at $268
\,$GHz and shows that, though there is a statistical correlation
between the atmospheric noise and sky emissivity, the variation in the
measured atmospheric stability at a single value of sky emissivity is
almost as large as the variation between different sky emissivities.
This is thought to arise from night to night variations in the power
spectrum of the atmospheric turbulence, along with variations in the
height and drift speed of the turbulent layer(s), and is not
predictable from models.  It is therefore not surprising that, though
the sky emissivity is higher in the blank sky data, the nod
mis-subtraction is less pronounced since long timescale fluctuations
in the sky emissivity do not necessarily track with its absolute
level.

\begin{figure*}[ht]
\centering
\epsfig{file=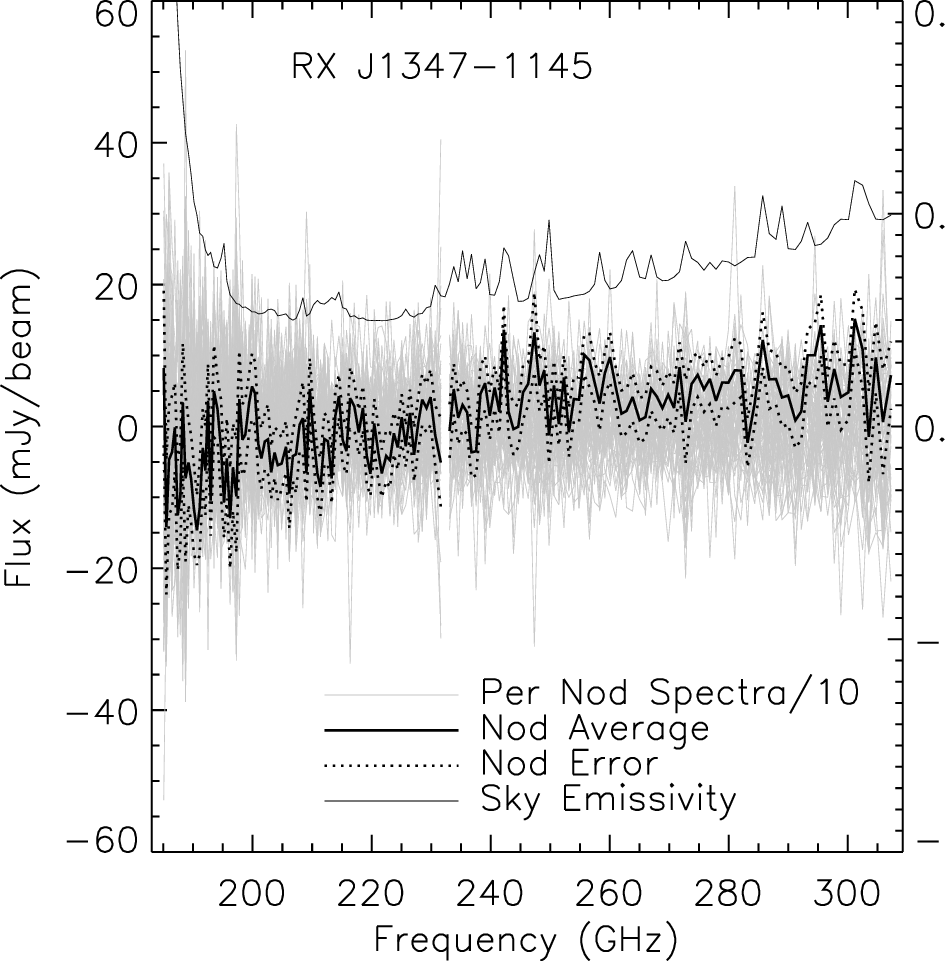,width=0.4\textwidth}
\hspace{1.2cm}
\epsfig{file=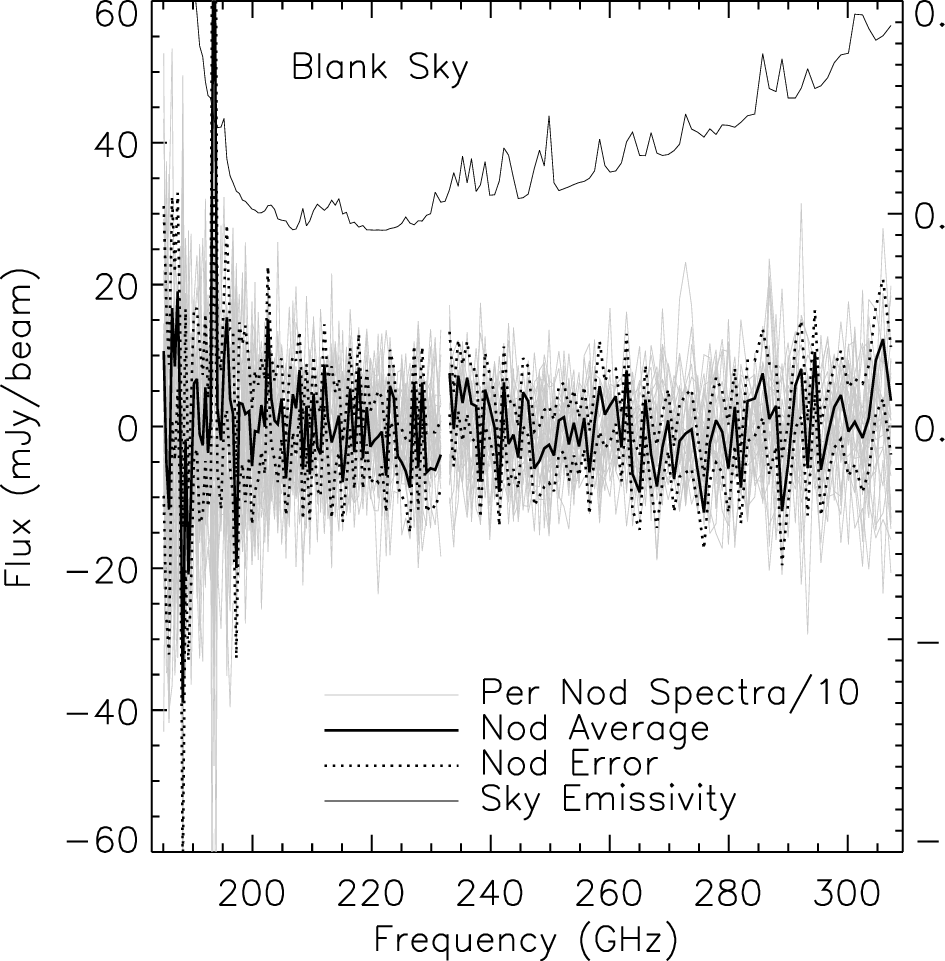,width=0.4\textwidth}
\caption{Nod spectra, averages and sky emissivity for the \rxj\ (left
  panel), and blank sky (right panel) field using only those data
  passing the data cuts constructed with the algorithm presented in
  Section \ref{sS:reduction}.  The data are presented as described in
  Figure \ref{fig:allnods}.  The masking procedure has rejected those
  nod data with spectra following the sky emission spectrum, leaving a
  flat spectrum in the case of the blank sky data.  When the same
  algorithm is used on the \rxj\ data, the left-hand panel results.}
\label{fig:cutnods}
\end{figure*}

The data cutting algorithm we develop is built on the assumption that
the continuum sky signal has very little curvature across the Z-Spec
band, while the atmosphere has the ``U''-shaped spectrum shown in Figure
\ref{fig:allnods}.  In detail, the algorithm employs the following
steps.  First, the noise-weighted spectral average of all nods with no
cut is generated and the line $\mathbf{y}$ is fit to it; as the
simplest model of a continuum astronomical signal the linear model
provides a template against which the measured spectra can be
compared.  Next, for each nod's spectrum $\mathbf{d}$, the quantity
\begin{equation}
\label{eq:nodcut}
\xi = \sum_{i=1}^{10} |\frac{d_{i} - y_{i}}{\sigma_{i}}| +
\sum_{i=150}^{160} |\frac{d_{i} - y_{i}}{\sigma_{i}}|,
\end{equation}
is computed, where $\mathbf{\sigma}$ is the uncertainty of the spectrum
and $i$ runs over the shortest 10 frequency bins in the first summand
and the longest 10 in the second.  The $\xi$ statistic is a measure
of how much the low- and high-frequency components of the spectra
deviate from the linear model, and essentially corresponds to a sum of
the number of sigma away from the mean these points lie.  The first
and last 10 frequency bins are chosen because these are the most
sensitive to the water lines bracketing the Z-Spec band and most
likely to exhibit failures in the nod modulation; empirically we find
that points further inside the band than this do not have much
curvature, though at the high-frequency end this is a relatively soft
choice.  To generate the mask, all nods with $\xi$ greater than some
value $\xi_{0}$ are flagged to be cut.  The process is then iterated
with the average and line fit being generated from the remaining nods
and a recomputation of $\xi$ for each nod.  The free parameters of
this algorithm are the number of iterations to perform and the value
of $\xi_{0}$: we empirically find that the blank sky spectrum is
consistent with zero when cuts are performed using two iterations of
the algorithm and $\xi_{0} = 10$, though the result is not strongly
sensitive to the choice of $\xi_{0}$ for $\xi_{0} \leq 20$.  Figure
\ref{fig:cutnods} shows the result of this procedure for the blank sky
data, which has a spectrum consistent with zero after the cuts are
applied.  When the same algorithm is performed on the \rxj\ data, the
average spectrum shown in the left hand panel of Figure
\ref{fig:cutnods} results.  This spectrum is now statistically
uncorrelated with the sky emissivity.

The fraction of data cut using this procedure is $31 \,$\% for
\rxj\ and $43 \,$\% for the blank sky field.  In the case of the
\rxj\ observations for which there are data from six continuous
nights, we note that the large proportion of the data cut by the
algorithm belong to only three of the observing nights (during which
only 27 \% of the total data volume was acquired).  This matches the
expectation that the level of atmospheric fluctuations are similar
over timescales of several hours, but can change substantially between
nights \citep{Sayers2010}.

\begin{figure*}[ht]
\centering
\epsfig{file=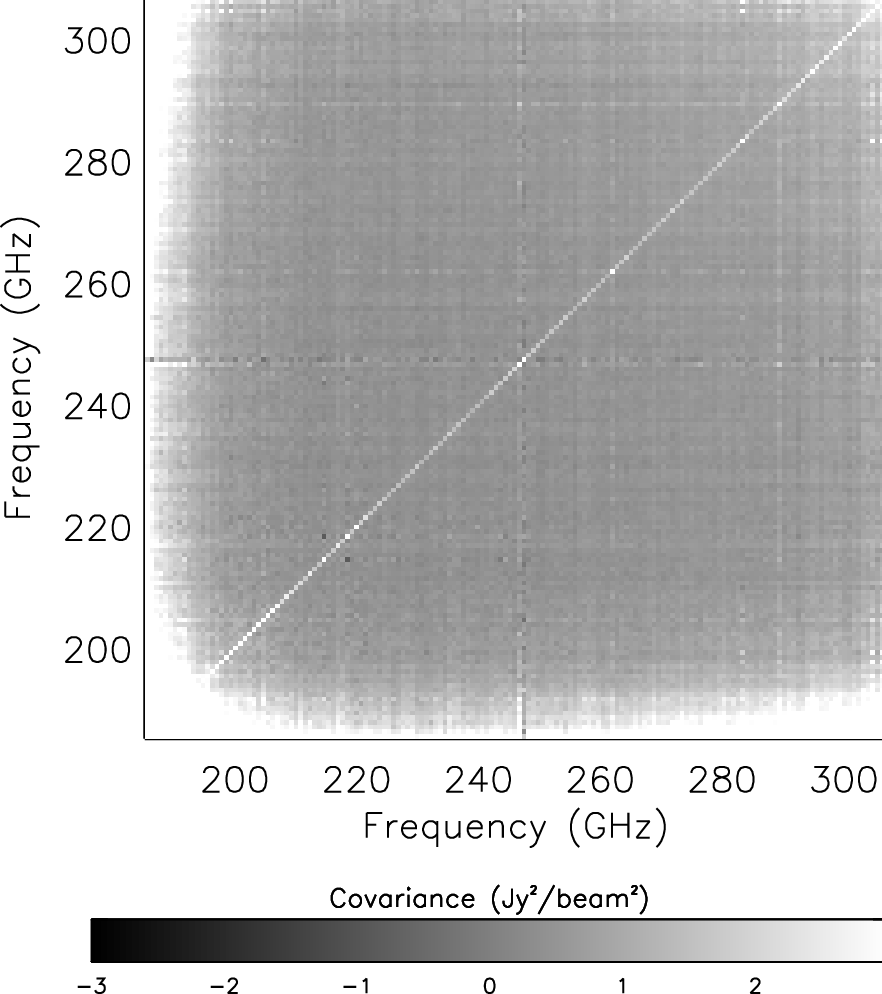,width=0.4\textwidth}
\hspace{1cm}
\epsfig{file=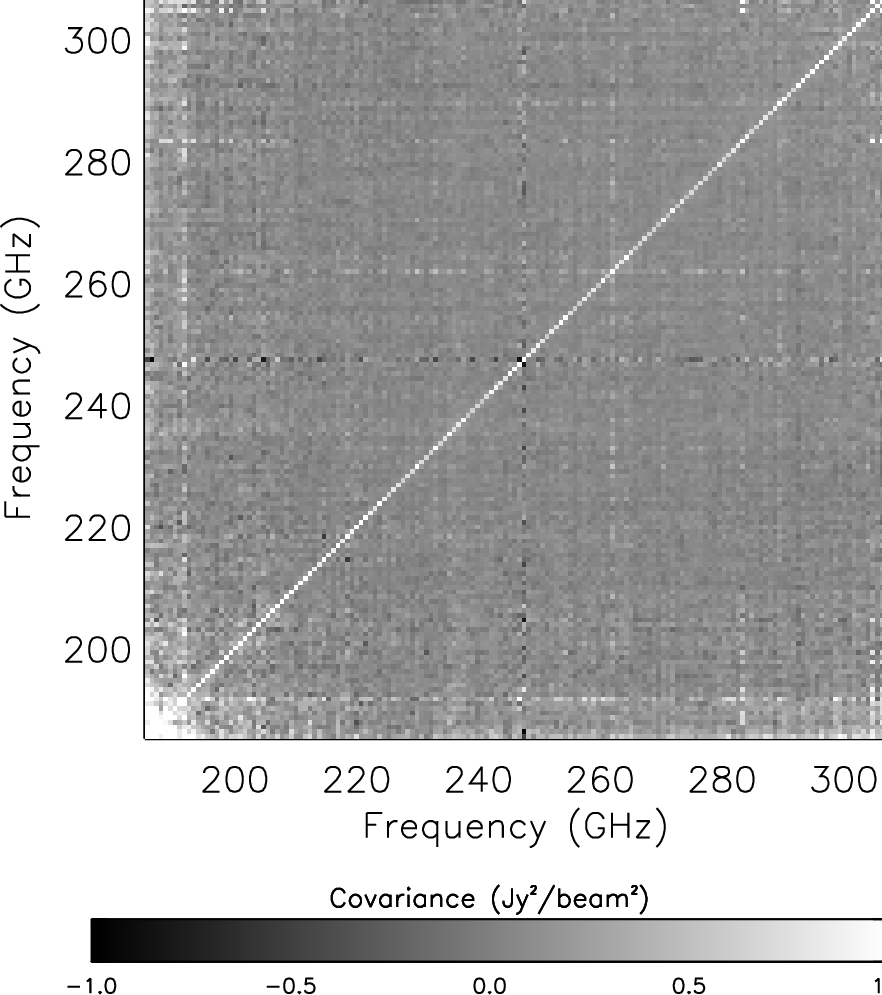,width=0.4\textwidth}
\caption{Covariance matrix of the full and cut nod data.  The
  left-hand panel shows the covariance of the full data set; the
  characteristic shape of the sky emissivity is present in the
  bolometer--bolometer covariance, suggesting that the correlations in
  the individual nod spectra arise from variations in the sky
  emissivity.  The right-hand panel shows the same covariance for the
  nod set with the data cuts described in Section \ref{sS:reduction}
  applied; the atmospheric shape has been removed by cutting those
  nods which have incompletely canceled the variations in the sky
  emission.}
\label{fig:covariance}
\end{figure*}

It is possible that there is covariance between bolometer time streams
in these data, arising from optical or electrical coupling between the
detectors, or correlations in the atmospheric emission itself.  To
check this, we compute the mean bolometer--bolometer covariance matrix
using
\begin{equation}
\mathcal{C} = \frac{1}{N_{\mathrm{nod}}-1} \sum_{i=1}^{N_{\mathrm{nod}}}
 \left( \mathbf{d_i} - \mathbf{\bar{d}} \right) 
 \left( \mathbf{d_i} - \mathbf{\bar{d}} \right)^{T},
\label{eq:covmat}
\end{equation}
where again $\mathbf{d_{i}}$ is the spectrum of a single nod,
$\mathbf{\bar{d}}$ is the average spectrum over all nods, and the
summand runs to the number of nods $N_{\mathrm{nod}}$.  The matrix
$\mathcal{C}$ has been calculated for both the full set and the cut
set of \rxj\ observations; these are shown in Figure
\ref{fig:covariance}.  These show that in the full set of nods the
covariance matrix is tracing out the structure of the atmosphere; the
signature ``U'' shape of the sky emissivity is visible along the parts
of the covariance matrix involving the lowest and highest frequency
bins.  Applying the data cuts largely removes these correlations to
the point that only the three lowest frequency bins remain strongly
covariant; these are excluded from further analysis.  We take this as
good evidence that residual covariance in the nod spectra arising from
short-term fluctuations in the sky emissivity is negligible in the
cut data set.

Since neighboring spectral channels are uncorrelated, Z-Spec can
measure the absolute level of the spectrum to an accuracy dictated by
the (independent) uncertainty in each channel.  To estimate the
uncertainty in the zero point of the Z-Spec spectra, we compute the
band average of a line fit to the blank field spectrum.  This
procedure yields an estimate of $\pm 0.2 \,$mJy on the zero point
which can applied to the spectra as a systematic error.

\begin{figure*}[ht]
\centering
\epsfig{file=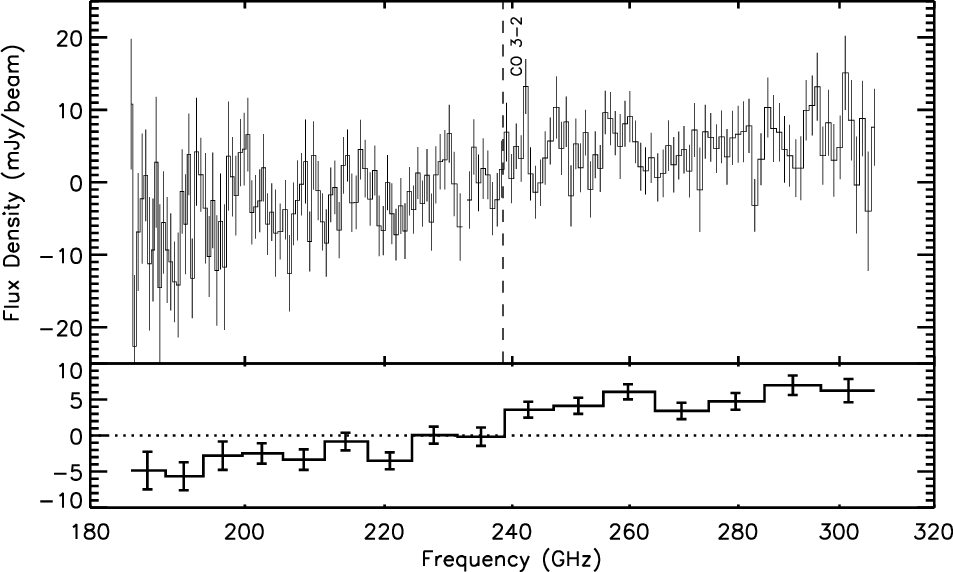,width=0.95\textwidth}
\caption{Measurement of the SZ effect in \rxj\ with Z-Spec's native
  binning (upper panel) and binned into 16 spectral element wide
  noise-weighted averages (lower panel).  Shown for reference is the
  position of the CO $J=3 \to 2$ line at the redshift of the galaxy
  cluster; this line would be the brightest available in a dusty
  sub-mm galaxy, were one coincident with the pointing of the
  measurement.  No individual lines are detected in this spectrum.
  The lower panel shows that the spectrum of this pointing has a
  decrement in flux at the lowest wavelengths and an increment at the
  highest, as would be expected from the SZ effect (see Figure
  \ref{fig:defaultSZspectrum}).}
\label{fig:measuredspectrum}
\end{figure*}

Once these cuts are applied to the data, the resulting spectrum is
shown in Figure \ref{fig:measuredspectrum}, both in the native Z-Spec
binning and as uncertainty-weighted averages in 16 native-element wide
bins.  Though the native resolution spectrum has low signal-to-noise
ratio per measurement, the average shows a clear decrement at low
frequencies and increment at high frequencies, which is a signature of
the SZ effect; this is investigated further in Section
\ref{S:results}.

\subsection{Presence of Line Emission}
\label{lines}

Emission from the SZ effect is purely continuum; if the emission
measured with Z-Spec is associated with the SZ effect, no lines will
be present in the spectrum.  Of course, contamination from sub-mm
galaxies along the same line of sight as the cluster could introduce
line emission in the Z-Spec spectrum, and further, the presence of
such galaxies could contribute (positive) continuum flux to the Z-Spec
measurement and bias the result.  It is important to search for the
presence of line emission in the Z-Spec measurement to show that no
contamination from sub-mm galaxies is present in the spectrum.

Current measurements show that two potentially sub-mm bright galaxies
reside in this cluster.  The central galaxy has been detected by
MUSTANG at $90 \,$GHz \citep{Mason2010} but is not detected by SCUBA
at $350 \,$GHz \citep{Zemcov2007}.  In the Z-Spec bands we calculate
$S < 2\,$mJy across the Z-Spec band for this galaxy (see Section
\ref{sS:simulations}); this flux will be further diluted by a factor
of $\sim 5$ due to its position in the Z-Spec beam.  A sub-mm bright
galaxy is known exist to the southwest of the cluster center at
$13^{\mathrm{h}} 47^{\mathrm{m}} 27^{\mathrm{s}}.6, -11^{\circ} 45'
54''$ (\citealt{Zemcov2007}; C.~D.~Dowell 2008, private
communication), but this source is far from the both the Z-Spec
pointing and chop positions (see Figure \ref{fig:RXJ1347_chop}).  We
are therefore confident that whatever sub-mm galaxy emission may be
present in the Z-Spec spectrum, it is not associated with known,
sub-mm bright sources.

No lines are individually detected in the spectrum shown in Figure
\ref{fig:measuredspectrum}.  To search for the presence of faint line
emission from sub-mm galaxies, we adopt the algorithm described in
\citet{Lupu2011} using the final \rxj\ spectrum.  The spectrum of
\rxj\ fails to pass any of the redshift determination criteria
described in \citet{Lupu2011}.  The significance of the redshift
determined using this algorithm is $18 \,$\%, if the noise between
channels is uncorrelated under the assumption that multiple lines are
present in the spectrum.  The spectrum does contain a single
$\sim3\sigma$ feature at $242 \,$GHz, however $3 \sigma$ deviation in
a single bin should arise naturally from statistics given the number
of Z-Spec channels.

Of course even in the absence of detected sub-mm sources, the faint,
confused sub-mm background will also be present in the Z-Spec beam.  A
detailed investigation of the effect of the continuum emission from
such sources is given in Section \ref{sS:systematics}; unfortunately,
it is difficult to place limits on the effect of line emission from
such sources on this measurement.  As a simple check, we compute the
histogram of the signal-to-noise ratios of the spectral bins after a
quadratic fit is subtracted.  As above, a simple quadratic model is a
good approximation to the chopped SZ spectral shape over the Z-Spec
passband, so is used to remove the continuum emission in the measured
spectrum.  This resulting histogram should have unity standard
deviation if the variation of the data is well described by the
instrumental noise estimates; in contrast, if faint lines are present
the distribution width will be larger.  We fit a Gaussian distribution
to the histogram of spectral signal-to-noise ratios and calculate a
standard deviation of $1.07 \pm 0.07$ for the distribution.  This is
consistent with the absence of faint lines in the spectrum to the
uncertainties of the measurement.

\vspace*{1cm}

\section{Results}
\label{S:results}

The spectrum shown in Figure \ref{fig:measuredspectrum}\ shows a
decrement and increment structure which is indicative of the SZ
effect, passing through a null near $220 \,$GHz as expected.  However,
the amplitude of this change is significantly smaller than expected
from the previously measured SZ effect (as exemplified in Figure
\ref{fig:defaultSZspectrum}).  Investigation shows that this reduction
is due to the chopping attenuation; Figure \ref{fig:RXJ1347_chop}
shows that the chop throw of $90$ arcsec does not fall off the cluster
but rather on a region measured by Bolocam to have $\Delta
T_{\mathrm{CMB}} \sim -400 \, \mu$K.  Since the peak SZ effect in this
cluster is measured to have $\Delta T_{\mathrm{CMB}} \sim 850 \,
\mu$K, we expect a reduction in the measured SZ amplitude of a factor
$\sim 2$ from the chop.

Though sensitive to errors in the zero point of the measurement, the
frequency of the null in the measured tSZ effect should be independent
of its amplitude.  As discussed in Section \ref{S:intro}, both the rSZ
and kSZ corrections can shift the frequency of the tSZ null, and the
presence of such a shift in these high spectral resolution Z-Spec
measurements would be indicative of the presence of these effects.
Though more complex treatments are presented below, in the first
instance we use the model which requires the fewest assumptions to
determine the shift from the nominal $\nu_{0}$ of the null, which is
that the Z-Spec measurements can be approximated by a linear model
near the tSZ zero crossing.  The null frequency is determined by
fitting a line to the Z-Spec measurements for $\nu < 260 \,$GHz, above
which the Z-Spec spectrum measurably deviates from a linear model.
The best-fitting model then yields a position for the null whose
uncertainties can be calculated from the covariance matrix of the fit
using standard error propagation.  This calculation using the Z-Spec
measurements yields a null frequency of $\nu_{0} = 225.8 \pm 2.5
\,$GHz, which is formally a $3.3 \sigma$ detection of the shift in the
null of the SZ effect from the canonical tSZ value of $217.5 \,$GHz
assuming statistical errors only.  If the $\pm 0.2 \,$mJy systematic
uncertainty of the zero point is included an additional uncertainty of
$\pm 1.2 \,$GHz is accrued, leading to a $3.0 \sigma$ measurement of
the null shift.

As presented in \citet{Itoh1998}, the shift in the null from $217
\,$GHz can be approximated by\footnote{Note that more complex
  relations involving other parameters can be formulated,
  e.g., \citet{Hansen2004}; here we use the simplest relation between
  $x_{0}$ and $\theta$ since the signal-to-noise ratio of the data
  does not support more parameters.}
\begin{equation}
x_{0} = 3.830 (1 + 1.1674 \theta_{\mathrm{e}} - 0.8533
\theta_{\mathrm{e}}^{2}),
\label{eq:itohshift}
\end{equation}
where $x_{0}$ is the dimensionless null frequency and
$\theta_{\mathrm{e}} = k_{\mathrm{B}} T_{\mathrm{e}} / m_{\mathrm{e}}
c^{2}$.  Under the assumption that $v_{\mathrm{pec}} = 0$, the null
frequency measured by Z-Spec implies $k T_{\mathrm{e}} = 17.1 \pm 5.3
\,$keV in \rxj; though not the first constraint on cluster temperature
from the SZ effect (\citealt{Hansen2002}; \citealt{Zemcov2010}), this
measurement constitutes the first direct measurement of the
temperature of a cluster using the shift in the SZ null frequency.
This temperature is consistent with the $k T_{\mathrm{e}} \sim 16 \,$keV
derived from X-ray data (Section \ref{sS:vpec}).

\subsection{SZ Likelihood Functions}
\label{sS:simulations}

The simplest way to correct the true amplitude of the SZ effect in
\rxj\ for the effects of Z-Spec's chop and demodulation losses is via
a simulation which includes a model of the cluster, the observation
scheme employed by Z-Spec, and the effects of the data analysis
pipeline.  

In this work, the cluster SZ shape is modeled using a combination of
the SZ effect parameters measured directly by Bolocam for the bulk SZ
component and by adopting the model of \citet{Mason2010} for the fine
scale component.  Since, per resolution element, Z-Spec is essentially
a single-pixel narrowband photometer, the spatial model of the
cluster cannot be constrained by these Z-Spec data but rather dictates
the ratio of the SZ signal in the on- and off-source beams given an SZ
effect amplitude (see, for example, Figure 1 of \citealt{Zemcov2003}
and the associated discussion).  In this work, the bulk SZ effect
model (reflecting the SZ emission on scales $\gtrsim 1'$) is derived
via a fit to an elliptical isothermal $\beta$ model using the Bolocam
$140 \,$GHz map of the cluster; the parameters used in the model are
given in Table \ref{tab:isothermalbeta}.  Uncertainties are given only
on the semimajor and semiminor core radii $(\theta_{\mathrm{c,min}},
\theta_{\mathrm{c,maj}})$, because these parameters are strongly
degenerate with changes in $\beta$ and the uncertainty on them
sufficiently encapsulates the uncertainty on the model shape.

The model for fine scale component (for emission on scales $10'' <
\theta < 1'$) we employ is developed in \citet{Mason2010}; since
Z-Spec's beam is large compared to the fine scale components, the
details of the fine scale component should not have much effect on the
SZ amplitude measured by Z-Spec (this is investigated further in
Section \ref{sS:systematics}).  The rSZ corrections derived by
\citet{Nozawa2000} are used to correct the tSZ signal for the effects
of the relativistic electron population.  As these numerical
corrections are only applicable for $k T_{\mathrm{e}} \leq 25 \,$keV,
all calculations in this work only investigate rSZ corrections up to
this temperature.

\begin{table}[t]
\centering
\caption{Isothermal $\beta$ model parameters for the bulk SZ
  emission.} 
\begin{tabular}{lc}
\hline
Bolocam $\nu_{0}$ & $140 \,$GHz \\ 
$y_{0}^{\mathrm{a}}$ (no rSZ correction) & $5.18 \times 10^{-4}$ \\ 
$\beta^{\mathrm{a}}$ & $0.86$ \\
$\theta_{\mathrm{core,min}}$ & $37.8 \pm 2.5 \,$arcsec \\
$\theta_{\mathrm{core,maj}}$ & $46.2 \pm 3.1 \,$arcsec \\
P.A. &  $-4.6^{\circ}$ \\
\hline 
\multicolumn{2}{l}{\textbf{Note.} $^{\mathrm{a}}$ See Section \ref{S:results} for a
  detailed discussion of the}\\ 
\multicolumn{2}{l}{uncertainties associated with these parameters.} \\ 
\end{tabular}
\label{tab:isothermalbeta}
\end{table}

In a given simulation, the bulk component of the cluster model is
given an amplitude corresponding to a particular set of the parameters
$y_{0}, T_{\mathrm{e},0},$ and $v_{\mathrm{pec,0}}$ reflecting the SZ
effect amplitude, ICM electron temperature, and peculiar velocity of
the cluster along the line of sight, respectively.  The fine scale
components are held at a constant $y_{\mathrm{clump}}$ and
$T_{\mathrm{e,clump}}$ and are included in the map.  Finally, the
central radio source in \rxj\ is added into the map with a flux given
by
\begin{equation}
\label{eq:cs}
S_{\mathrm{CS}}(\nu) = S_{\mathrm{CS},0} (\nu/\nu_{0})^{\alpha},
\end{equation}
where $S_{\mathrm{CS},0} = 60.1^{+49.5}_{-27.1} \,$mJy, $\nu_{0}$ is
fixed at $1 \,$GHz, and $\alpha = 0.77 \pm 0.21$; these parameters are
determined from an uncertainty-weighted fit to the measurements of
\citet{Condon1998}, \citet{Gitti2007}, \citet{Coble2007},
\cite{Cohen2007}, and \citet{Mason2010}.  Due to the frequency at which
Z-Spec operates and the offset between the Z-Spec pointing and
location of the central cluster source, the details of the source do
not have a large effect on the SZ measurement (see Section
\ref{sS:systematics}).  This model map is then smoothed with the
Z-Spec optical response function (ORF) at each wavelength; the ORF is
modeled by a Gaussian whose width varies according to
\begin{equation}
\label{eq:fwhm}
\mathrm{FWHM} = (a - b / \nu) / \nu,
\end{equation}
where the constants $a=8715''$ GHz and $b = 9312''$ GHz$^{2}$ are
empirically determined from dedicated beam mapping observations.

These simulated, ORF-convolved maps are then sampled in the same way
as Z-Spec sampled the true cluster to generate simulated time-ordered
data (sTOD).  These sTOD are then propagated through the Z-Spec
analysis pipeline starting at the chop demodulation stage using the
same settings, data cuts, and filtering as the real data and propagated
to an equivalent simulated spectrum.  These simulations are noiseless,
that is, we do not develop a noise model for the instrument and
include random realizations of the instrument noise since we are not
estimating the instrumental errors on the spectrum via simulations in
this work.

A subtle attenuation factor due to the demodulation model arises from
the use of an artificial pure tone to demodulate the chop; this model
necessarily picks out only the component of the signal which occurs at
the first harmonic of $f_{0}$.  This means that, since signal power is
mixed to higher harmonics, some fraction of the signal is lost in the
$f_{0}$ estimate.  With real sky data the absolute calibration and
cluster observations are both observed and processed using the same
pipeline, so this attenuation is corrected during the calibration
step.  However, since the simulations are performed in sky-calibrated
units, this demodulation attenuation factor is not automatically
modeled by the simulations.  To account for this, we perform a
calibration simulation by populating an empty sky with a single unit
flux point source; this is a good model for a Z-Spec calibration
observation.  This simulated calibration map is then observed in the
same way as in a Z-Spec calibration observation.  The time streams are
then passed through the Z-Spec pipeline in the same way as the cluster
simulations to measure the attenuation of the flux due to the
demodulation model.  The resulting calibration correction factors,
which vary over the Z-Spec band from $\sim 50 \,$\% to $\sim 60 \,$\%
due to the beamwidth at the different Z-Spec resolution bins, are
then applied to the simulated spectra to account for the demodulation
model attenuation factor.  These corrected spectra can then be
compared directly to the measurements to compute a likelihood function
which varies as a function of the input parameters.

\begin{figure}[ht]
\centering
\epsfig{file=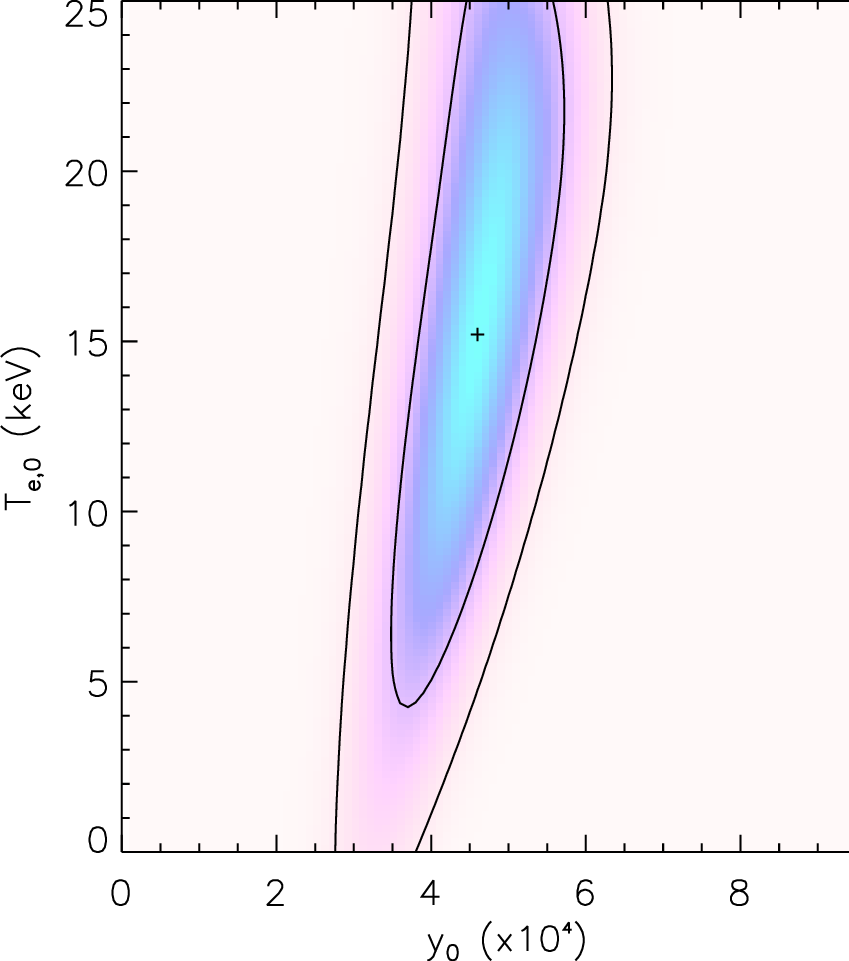,width=0.38\textwidth}
\caption{Likelihood function for the Z-Spec measurement using the
  cluster model described in Section \ref{sS:simulations} and assuming
  $v_{\mathrm{pec}}=0$.  The color is proportional to the likelihood
  and the contours show the $68\,$\% and $95 \,$\% confidence regions.
  The Z-Spec data have a peak likelihood for a model with $y_{0} = 4.6
  \times 10^{-4}$ and $k T_{\mathrm{e}} = 15.2 \,$keV with a large
  uncertainty on the bulk cluster temperature.}
\label{fig:lh_zspeconly}
\end{figure}

The most likely cluster parameters are determined using the likelihood
function based on this simulation pipeline and the real measurements.
The simulator is run with a variety of $y_{0}, T_{\mathrm{e,0}}$, and
$v_{\mathrm{pec}}$; for each parameter set, the $\chi^{2}$ statistic
is computed from the true spectrum measurements and their
uncertainties with the simulated spectrum used as the model.  The
likelihood function is then formed from the grid of $\chi^{2}$ values
in the standard way.  Figure \ref{fig:lh_zspeconly} shows the
resulting likelihood in the case where $v_{\mathrm{pec}}=0$, i.e., only
$y_{0}$ and $T_{\mathrm{e},0}$ are varied.  As shown in the figure,
the maximum likelihood occurs at $y_{0} = 4.6^{+0.6}_{-0.9} \times
10^{-4}$ and $T_{\mathrm{e},0} = 15.2^{+12}_{-7.4} \,$keV, where the
uncertainties represent the marginalized $68 \,$\% confidence
intervals of the likelihood.\footnote{Note that in the case of
  $T_{\mathrm{e},0}$ we are unable to place a reliable upper bound due
  to the lack of rSZ corrections at these very high temperatures.  We
  therefore adopt a conservative upper limit derived from a reasonable
  extrapolation of the marginalized likelihood curve.}  This means
that the hypothesis that no rSZ corrections are present is rejected at
the $2 \sigma$ level, which is less significant than the result from
the simpler analysis presented in Section \ref{S:results}.

\begin{figure}[ht]
\centering
\epsfig{file=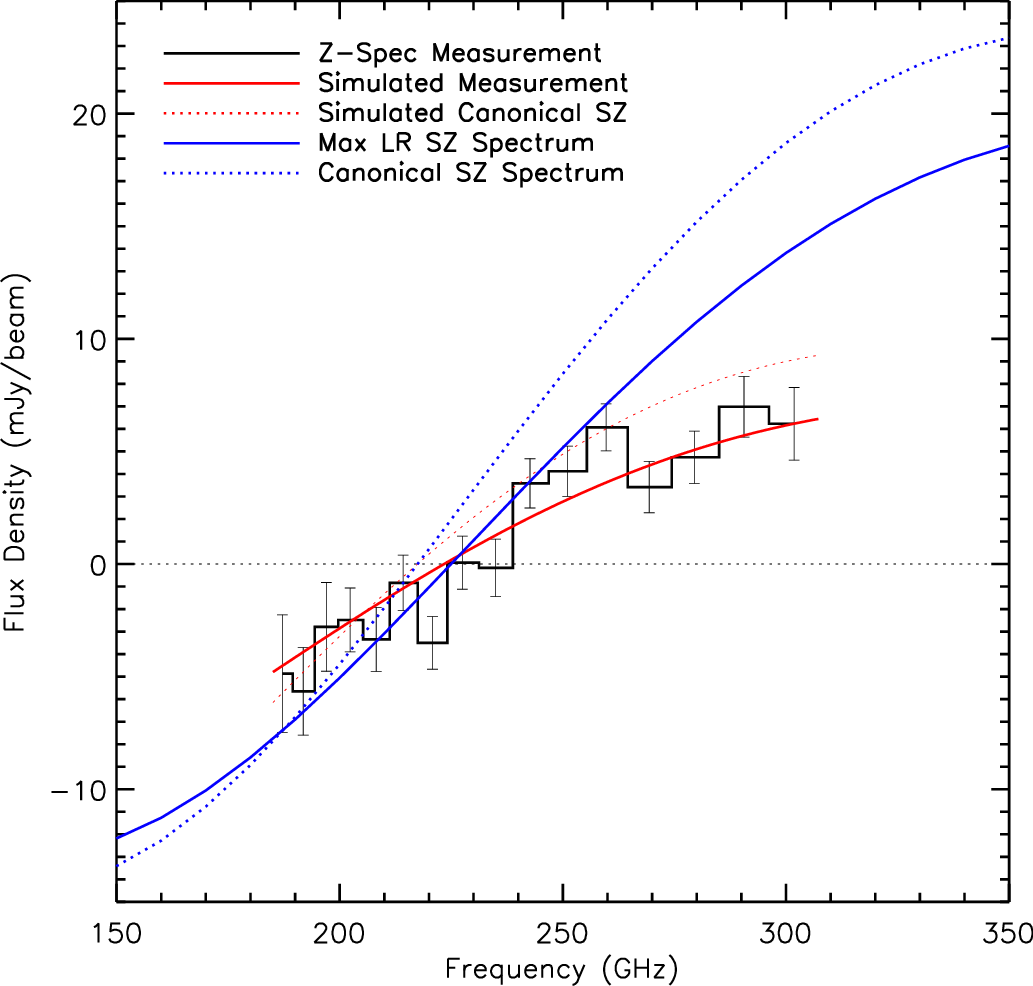,width=0.43\textwidth}
\caption{SZ effect distortions corresponding to the most likely models
  given the Z-Spec spectral measurement of \rxj.  The black line shows
  the measured Z-Spec points, while the red shows the minimum
  $\chi^{2}$ model matching the data arising from the input shown by
  the solid blue line.  The difference between the red and blue lines
  is largely due to the effects of the chop and nod observation
  strategy and to a lesser extent the varying beams over the Z-Spec
  bandpass.  The dotted blue line shows the canonical tSZ effect
  spectrum corresponding to the peak in the likelihood function, and
  dotted blue shows this canonical spectrum propagated through the
  Z-Spec pipeline; we can differentiate between the canonical and rSZ
  corrected models using their different amplitudes and shapes.}
\label{fig:meassim}
\end{figure}

\begin{figure}[ht]
\centering
\epsfig{file=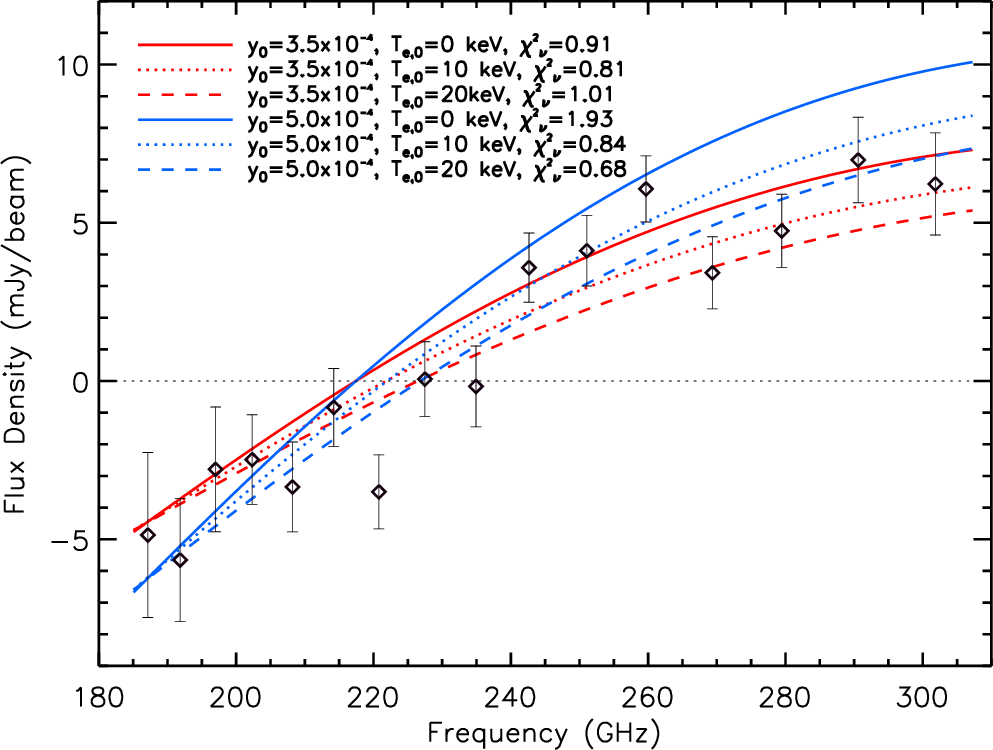,width=0.44\textwidth}
\caption{Difficulty in constraining $T_{\mathrm{e,0}}$ is caused
  by the combination of Z-Spec's relatively narrow free spectral range
  and the small amplitude of the shift in the SZ null from the rSZ
  effect.  This plot shows the binned Z-Spec measurements as the black
  points, and families of constant $y_{0}$ as red and blue lines.
  Different values of $T_{\mathrm{e,0}}$ are shown using the line
  styles as given in the legend, which also lists the $\chi^{2}$ of
  the fits to the Z-Spec data.  As can be seen in this figure, over
  the Z-Spec bandpass $y_{0}$ dictates the amplitude of the SZ effect,
  while increasing $T_{\mathrm{e,0}}$ ``smooths'' the SZ spectrum to
  shorter $\nu$, shifting the null slightly.  This means that the
  likelihood for a low $y_{0},T_{\mathrm{e,0}}$ combination can be
  similar to that for a higher $y_{0},T_{\mathrm{e,0}}$ pair.  This,
  in turn, leads to the shape of the likelihood function shown in
  Figure \ref{fig:lh_zspeconly}.}
\label{fig:Te_degen}
\end{figure}

Figure \ref{fig:meassim} compares the measured Z-Spec spectrum with
the simulated output and model corresponding to the maximum of the
likelihood function.  This figure highlights the effects of the chop
pattern attenuation on the measured signal; for example, at $300
\,$GHz the brightness of the measured emission is $6 \,$mJy
beam$^{-1}$, though the simulator allows us to infer that the true SZ
brightness should be $13 \,$mJy beam$^{-1}$ at this frequency in the
absence of the chop.  Figure \ref{fig:meassim} highlights a
shortcoming of Z-Spec for SZ measurements, which is that the
instrument's effective passband is too narrow to effectively constrain
the curvature of the SZ effect spectrum over the band.  In terms of
the parameters of the spectral model, because $y_{0}$ dictates the
amplitude of the signal over the Z-Spec bandpass and the SZ spectrum
is close to linear over it, the constraints on this parameter from
Z-Spec alone are reasonable given the instrumental sensitivity.
However, since both the rSZ and kSZ signals are spectrally broad
modifications to the color of the SZ effect, they are much less
constrained by even relatively wideband measurements near the SZ
null.  The case of the rSZ corrections is exemplified in Figure
\ref{fig:Te_degen}: over the Z-Spec bandpass (which covers the entire
$220 \,$GHz atmospheric window), the rSZ corrections predominantly
have the effect of flattening the spectrum of the SZ increment.  This
means that, in a $\Delta \chi^{2}$ sense, a measurement of
$T_{\mathrm{e},0}$ is largely degenerate with a different value of
$y_{0}$.  Further, because the rSZ corrections are so small, the
$\Delta \chi^{2}$ accrued from even large changes in
$T_{\mathrm{e},0}$ are not highly constraining.  This is what leads to
the poor constraints on $T_{\mathrm{e},0}$ in Figure
\ref{fig:lh_zspeconly}; clearly, even high-resolution measurements of
the SZ null are not very constraining on the combination of SZ
spectral corrections when taken over a relatively narrow bandpass.
The situation is even worse in the case of kSZ, which will be
discussed in Section \ref{sS:vpec}.

\subsection{Modeling Uncertainties}
\label{sS:systematics}

The various uncertainties in our model of the cluster emission will
lead to uncertainties in the SZ effect parameters measured using the
likelihood function; these can be estimated by measuring the variation
in the output SZ parameters caused by changes in the model input to
the simulator.  To do this, the input model parameters are varied
within their individual uncertainties and shifts in the best-fitting
results are measured.  Table \ref{tab:systematics} lists the model
parameters which are varied and the shift each causes from the result
$y_{0} = 4.6 \times 10^{-4}, T_{\mathrm{e,0}} = 15.2 \,$keV presented
in Section \ref{sS:simulations}.  

\begin{table*}[ht]
\centering
\caption{Systematic uncertainties in Z-Spec SZ measurement.}
\begin{tabular}{lccc}
\hline
Systematic Uncertainty & Parameter Variation & $\delta y_{0}$ ($\times
10^{4}$) & $\delta T_{\mathrm{e},0}$ (keV) \\ \hline

Bulk cluster shape & $\theta_{\mathrm{c}} + \sigma_{\theta}$ & $+0.2$ & $0$ \\

& $\theta_{\mathrm{c}} - \sigma_{\theta}$ & $-0.1$ & $+0.8$ \\

Shock region & $1.5 \times y_{\mathrm{clump}}$ & $-0.2$ & $0$
\\

 & $0.5 \times y_{\mathrm{clump}}$ & $+0.2$ & $+0.2$ \\

 & $0.5 \times T_{\mathrm{e,clump}}$ & $0$ &
$+0.8$ \\ 

 & Off & $+0.5$ & $-0.2$ \\

Central source & $S_{\mathrm{CS}}+\sigma_{S}$ & $-0.1$ & $0$ \\

 & $S_{\mathrm{CS}}-\sigma_{S}$ & $+0.2$ & $+0.4$ \\

 & Off & $+0.2$ & $+1.6$ \\

CIB & \citet{Bethermin2011} model & $0.0$ & $\pm 0.1$  \\ \hline

Absolute calibration & Abs.~Cal.~$+10 \,$\% & $+0.5$ & $-0.2$ \\ 

 & Abs.~Cal.~$-10 \,$\% & $-0.5$ & $+0.2$ \\ 

\hline
\end{tabular}
\label{tab:systematics}
\end{table*}

To estimate the effect of the uncertainties in the shape of the bulk
cluster emission, we vary the isothermal $\beta$ model to the $1
\sigma$ uncertainties in the Bolocam measurement of the ICM shape as
listed in Table \ref{tab:isothermalbeta}.  Note that we have fixed
$\beta = 0.86$ in the ICM model fits to the Bolocam data due to the
strong degeneracy between $\beta$ and $\theta_{c}$.  The degeneracy is
especially strong over the relatively modest spatial dynamic range of
both the Bolocam and Z-Spec data, and therefore fixing the value of
$\beta$ does not overconstrain the allowed range of ICM profiles.

The parameters of the proposed small-scale shocks in \rxj\ are not
well understood; it is necessary to investigate the effect of
misestimating their amplitudes on the Z-Spec result.  The Z-Spec
observations are centered on the shock region to the southeast of the
cluster center; we therefore expect this shock to have a much greater
effect on the Z-Spec results than the smaller shock to the east of the
cluster.  Because of this we co-vary the two shock parameters by the
amounts listed in Table \ref{tab:systematics} to both $150 \,$\% and
$50 \,$\% of their nominal $y_{0}$.  In addition, the shock
temperatures are reduced to $50 \,$\% of their nominal value of $25
\,$keV; since the rSZ corrections of \citet{Nozawa2000} are not
applicable at temperatures above this, we do not perform a simulation
where $T_{\mathrm{e,0}}$ is larger than its nominal value.  Finally,
to determine the overall effect of the shock regions, we perform a
simulation set where they are not present.

A further test is performed in which the central active galactic
nucleus (AGN) in the cluster is both varied by the $1 \sigma$
uncertainties on the spectral extrapolation to the Z-Spec band as
listed in Section \ref{sS:simulations}.  Additionally, a test is
performed where the AGN source is absent from the cluster model; the
results are again listed in Table \ref{tab:systematics}.

\begin{figure}[ht]
\centering
\epsfig{file=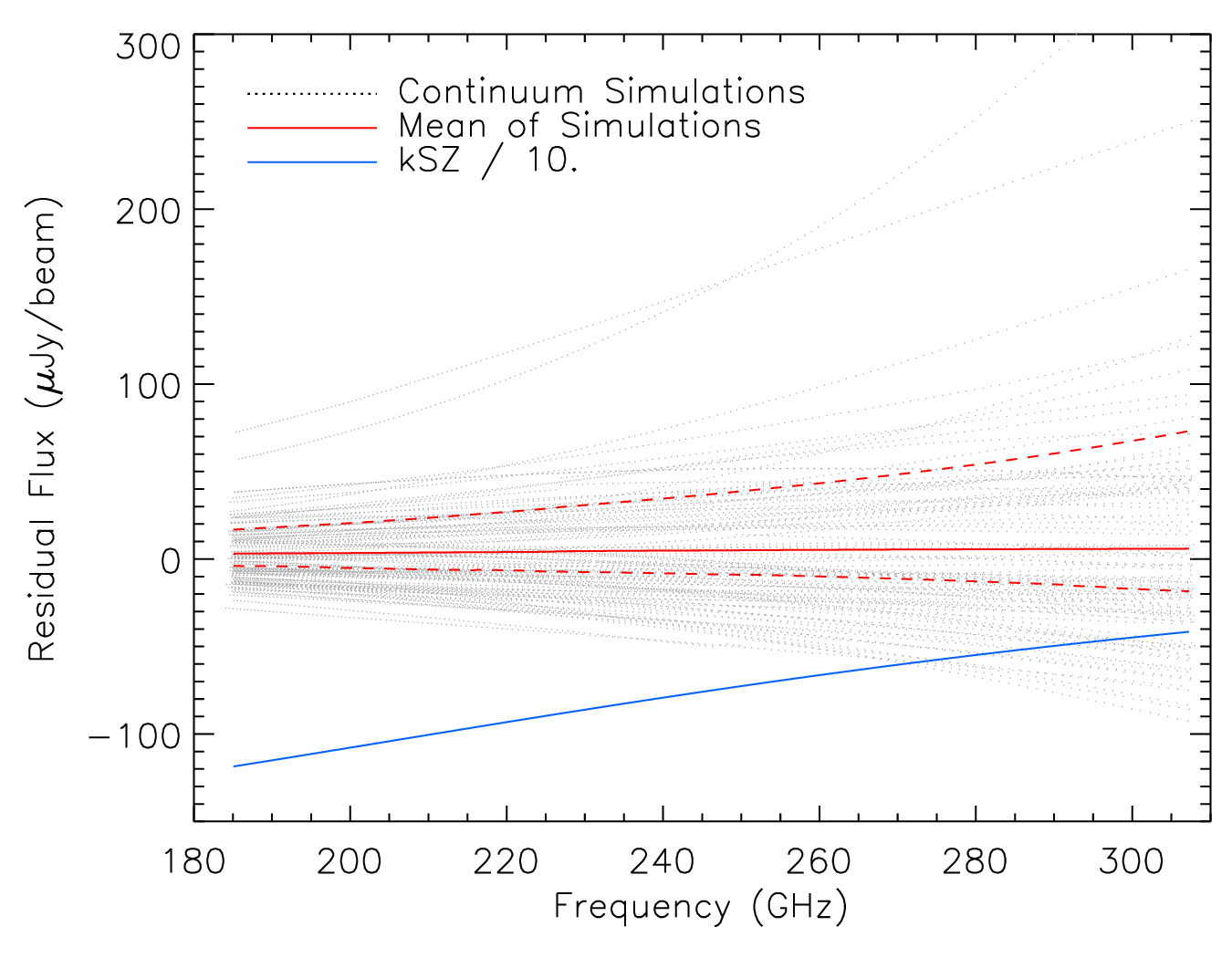,width=0.44\textwidth}
\caption{Simulations of the systematic effect of the faint sub-mm
  background on these SZ measurements.  Simulated maps generated using
  the \citet{Bethermin2011} CIB model are sampled and propagated
  through the Z-Spec pipeline, and the resulting spectra for 100
  random realizations are shown as dotted gray lines.  The mean and
  single-sided standard deviation of these simulated spectra are shown
  in red solid and dashed lines.  Finally, for comparison, $1/10$ of
  the amplitude of the best-fitting kSZ spectrum derived in Section
  \ref{sS:vpec} is plotted in blue.  The simulated spectra are
  asymmetric with respect to zero due to the Z-Spec chopping
  observation scheme.  Though the mean of these 100 simulations is
  very close to zero, the variation between them can be large, and can
  involve biases as large as $300 \, \mu$Jy if bright sub-mm sources
  are located in the main chopped beam.}
\label{fig:continuum}
\end{figure}

The cosmic infrared background (CIB) is potentially a large
contaminant to this SZ measurement, though Z-Spec's large beam size
and the steeply falling sub-mm spectra of these sources work in our
favor.  To determine the effect of the CIB on this measurement, we
employ a model for the sub-mm background and simulate Z-Spec spectra
which would result from random observations of it (i.e., analogs of
the control field discussed in Section \ref{sS:reduction}).  The CIB
model of \citet{Bethermin2011} is used to generate realizations of the
sky which have random spectral energy distributions for sources such
that each individual Z-Spec resolution element is properly correlated
with the others.  These maps are then sampled with the Z-Spec chop
pattern and the resulting time streams are passed through the Z-Spec
analysis pipeline and propagated to spectra as in the previous
simulations.  Figure \ref{fig:continuum}\ shows the resulting spectra
from 100 simulations of the sub-mm background.  The mean spectrum has
a positive bias but a mean of only a few $\mu$Jy beam$^{-1}$, and the
variance of the spectra has a complex spectral shape but is everywhere
$< 50 \, \mu$Jy beam$^{1}$.  This is approximately $5 \,$\% of the
amplitudes of either the relativistic corrections or kSZ effect in
this cluster; parameter constraints are listed in Table
\ref{tab:systematics}.

The results of these simulations highlight some interesting behavior
in the spectra when Z-Spec observes a random patch of sky.  First, as
expected, the variance due to the sub-mm background is larger at the
shorter wavelengths, about $50 \, \mu$Jy beam$^{-1}$ with some
realizations reaching as much as a few hundred $\mu$Jy beam$^{-1}$,
compared to the long Z-Spec wavelengths where it is $\sim 10 \,
\mu$Jy.  In addition, due to the chop, the spectra of the confused
sub-mm background is asymmetric with respect to zero flux.  This is
because the chop throw spreads the negative chop lobes over the sky,
diluting the effects of bright sub-mm sources, while the on-source
position is static on the sky.  This means that the rare occasions
when a bright sub-mm source\footnote{Note that here ``bright'' means
  $\sim 1 \,$mJy at $150 \,$GHz.} falls under the main lobe have large
contributions to the variance, while bright sources in the chop lobes
do not (see \citealt{Zemcov2003} for a detailed discussion).  We note
that these simulations do not include the effects of lensing, which
would amplify background sources near the center of clusters, thereby
increasing this bias.  Fortunately, in \rxj\ we are confident that the
sub-mm background is well measured and understood, and that no bright
sub-mm sources lie in either the on source or chopped positions (e.g.,
\citealt{Zemcov2007}; M. Zemcov et al.~in preparation).  However, in a
hypothetical large survey of clusters searching for spectral
corrections with a chopping instrument like Z-Spec, we expect that
this asymmetry bias due to the lensing amplification would be among
the most important systematic uncertainties.

In addition to overall SZ uncertainties arising from misestimates in
the cluster model parameters, we investigate the effect of the
uncertainty in the absolute calibration of Z-Spec on the SZ results.
Because they can be calibrated from the atmosphere or very bright
astronomical sources with known spectra, the relative gains of the
spectral channels themselves are well understood and produce a
negligible uncertainty on the final SZ results.  Specifically, the
channel to channel calibration uncertainties are a few percent,
leading to a $\sim 0.1 \,$\% uncertainty on the resulting SZ
amplitude.  However, there is a $\sim 10 \,$\% uncertainty in the
overall Z-Spec absolute calibration which could affect the scaling of
our results to others in the literature.  This absolute calibration
uncertainty reflects not only the statistical uncertainty of the flux
calibration but also the estimated uncertainties arising from the data
reduction chain; because of this, we do not vary low level parameters
which appear in the reduction pipeline separately as we do for the
cluster model parameters.  To estimate the effect of the absolute
calibration uncertainty, the calibration of the Z-Spec spectrum is
varied by $\pm 10\,$\% and simulation sets are performed.

As can be seen in Table \ref{tab:systematics}, the largest single
uncertainty in the SZ results arises from the presence of the central
source.  Of course, since it is a well-understood radio source
removing it entirely is a very conservative estimate, but this test is
illustrative of the effect the AGN source has on the inferred SZ
parameters.  The bulk cluster shape has the next largest effect on the
inferred $T_{\mathrm{e,0}}$, but in the context of the error on the
central value of $T_{\mathrm{e,0}}$, which is approximately $\pm 11
\,$keV, shifts at the $< 1 \,$keV level are essentially negligible.
The temperature of the shock region has a surprisingly large effect on
the SZ effect parameters, but again this is small compared to the
overall uncertainty on the ICM temperature.  Finally, the absolute
calibration has a roughly linear effect on the inferred $y_{0}$ but
little effect on $T_{\mathrm{e,0}}$; this is expected since the effect
of changes in $y_{0}$ directly tracks scalings in the absolute
calibration.

\subsection{Constraints on $v_{\mathrm{pec}}$}
\label{sS:vpec}

As a final study, we investigate the effect of the kSZ effect on the
Z-Spec measurements and the possibility of constraining it using these
data.  As shown in Figure \ref{fig:defaultSZspectrum}, the kSZ
contribution to the overall SZ spectrum has essentially no curvature
over the Z-Spec band; this is highlighted in Figure \ref{fig:kSZ}
below.  This means that any measurement of kSZ using these high
spectral resolution data requires tight constraints on possible
systematic errors in the zero point of the Z-Spec spectra.  This is
further compounded by the degeneracy between kSZ and rSZ if only the
shift in the null of the SZ effect is considered.  Thus far we have
only considered rSZ corrections in this work as rSZ has curvature over
the SZ band and its measurement does not rely on precise knowledge of
the spectrum zero point alone; the addition of kSZ changes the
constraining power of the data entirely.

\begin{figure}[ht]
\centering
\epsfig{file=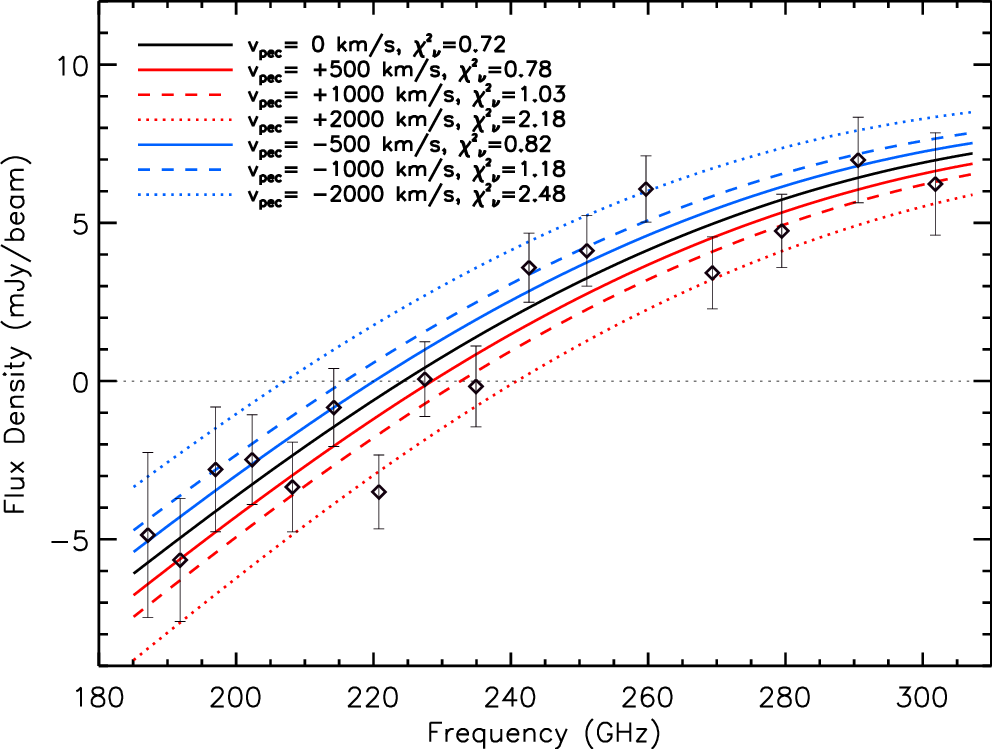,width=0.45\textwidth}
\caption{Effect of kSZ on the total SZ effect spectrum over the
  Z-Spec band.  The black points show the binned Z-Spec measurements
  and the colored lines families of SZ spectra for positive and
  negative values of $v_{\mathrm{pec}}$ assuming the best-fitting
  values of $y_{0}$ and $T_{\mathrm{e},0}$ given in Section
  \ref{sS:simulations}.  Different line styles are for different
  values of $v_{\mathrm{pec}}$, and the $\chi^{2}_{\nu}$ associated
  with each parameter set is listed in the legend.  Including the kSZ
  effect is degenerate with the addition of a zero point offset error
  to the measured spectrum and is thus difficult to constrain with
  such a narrow free spectral range.  However, large values of
  $v_{\mathrm{pec}}$ can be excluded by the Z-Spec measurements as
  knowledge of the zero point is secure to $\pm 1 \,$mJy beam$^{-1}$.}
\label{fig:kSZ}
\end{figure}

To help break these degeneracies and to reduce the range of
$\{y_{0},T_{\mathrm{e,0}},v_{\mathrm{pec}} \}$ space over which we
need to constrain the SZ spectrum, we rely on ancillary data.  The
Bolocam measurement of $y_{0}$ is extremely constraining since it is
limited by the absolute flux calibration of Bolocam, estimated as $\pm
5 \,$\% \citep{sayers11prep}.  For the canonical tSZ spectrum,
this corresponds to $y_{0} = (5.18 \pm 0.26) \times 10^{-4}$; because
Bolocam's band is in the SZ decrement, including a non-zero ICM
temperature simply increases this according to \citep{Itoh1998}
\begin{equation}
\label{eq:yt}
\frac{y_{0}(T_{\mathrm{e,0}})}{y_{0}} \approx \left(1-
\frac{T_{\mathrm{e,0}}}{150 \mathrm{keV}}\right)^{-1}. 
\end{equation}

In addition, $T_{\mathrm{e,0}}$ can be constrained using X-ray
measurements of the cluster.  We use {\it Chandra} observations of the
cluster which were carried out using the Advanced CCD Imaging
Spectrometer between 2000 and 2003.  The data reduction and
thermodynamic map creation follow the methods of \citet{Million2010}.
To determine the likelihood function for $T_{\mathrm{e,0}}$ from these
X-ray maps, we smooth both the temperature map and the temperature map
varied by the $1 \sigma$ temperature error map high and low with the
Z-Spec beam.  The temperature and its $1 \sigma$ limits can then be
measured at the point Z-Spec observed.  This procedure produces a
Gaussian likelihood function in $k T_{\mathrm{e}}$ with mean $16.4
\,$keV and standard deviation $1.8 \,$keV.

To generate the Z-Spec likelihood function, the Z-Spec simulator is
run over the $\{y_{0},T_{\mathrm{e},0}\}$ parameter space with
$v_{\mathrm{pec}}$ allowed to vary and using the simplifying
assumption $\tau_{\mathrm{e}} \approx y \, m_{\mathrm{e}} c^{2} /
k T_{\mathrm{e}}$ \citep{Birkinshaw1999}.  The likelihood space for
$\{y_{0},T_{\mathrm{e},0}\}$ for each $v_{\mathrm{pec}}$ is computed
in the standard way; $v_{\mathrm{pec}}$ is not tightly constrained in
a given simulation.

\begin{figure}[ht]
\centering
\epsfig{file=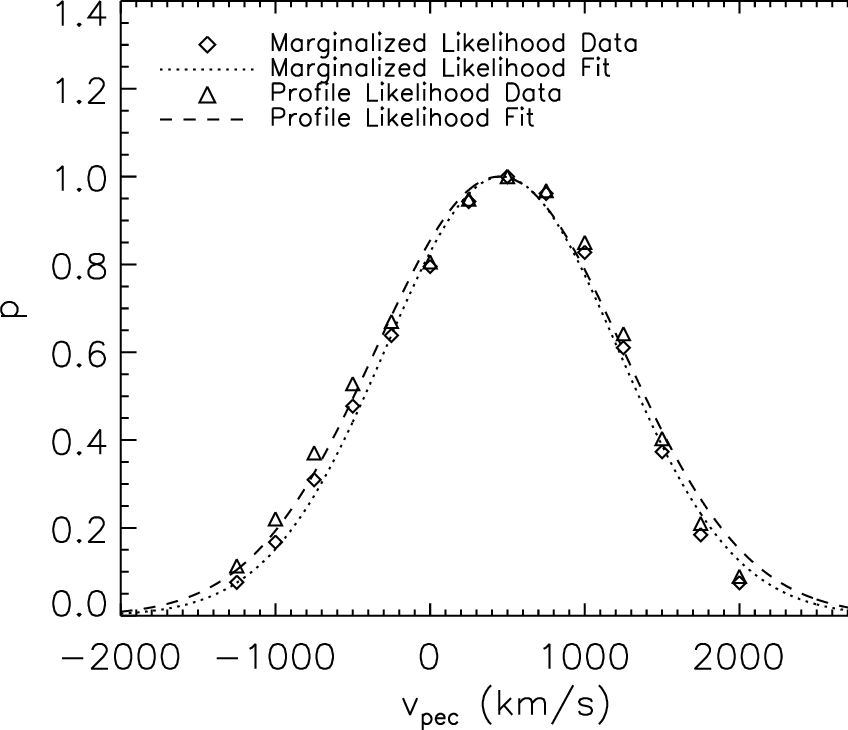,width=0.35\textwidth}
\caption{Joint marginalized and profile likelihood functions for
  $v_{\mathrm{pec}}$ in \rxj\ using Z-Spec, Bolocam, and X-ray data.
  The likelihood measurements are shown in points, and Gaussian
  distribution fits to these are lines as shown in the legend.  The
  maximum of the likelihood function yields the most probable
  $v_{\mathrm{pec}}$ and the width yields the statistical uncertainty
  on the measurement.}
\label{fig:vpecml}
\end{figure}

To constrain $v_{\mathrm{pec}}$, the joint likelihood function of the
Z-Spec, Bolocam and X-ray parameter sets is constructed by multiplying
the individual likelihoods together.  We use two methods to determine
the most likely $v_{\mathrm{pec}}$ and its uncertainty.  The first is
to marginalize over the parameters $y_{0}$ and $T_{\mathrm{e},0}$ in
the standard way, and the second is to compute the profile likelihood
function by finding the maximum likelihood of the solution
$\{y_{0},T_{\mathrm{e},0}\}$ pair for each $v_{\mathrm{pec}}$.  These
should be close to identical if the likelihood function is well
behaved.  Figure \ref{fig:vpecml} shows the resulting marginalized and
profile likelihood functions for $v_{\mathrm{pec}}$.  Since these
distributions should be well described by a Gaussian function, we fit
Gaussians to determine the maximum likelihood and confidence interval
on $v_{\mathrm{pec}}$, leading to an estimate of $v_{\mathrm{pec}} =
{+} 450 \pm 800 \,$km s$^{-1}$ for \rxj\ from these combined data
sets.  The simulations do not include an estimate of the effect of the
uncertainty on the absolute offset of the Z-Spec spectrum, which given
the uncertainty estimate of $\pm 0.2 \,$mJy beam$^{-1}$ equates to a
$v_{\mathrm{pec}}$ uncertainty of $\pm 110 \,$km s$^{-1}$.  Adding
this uncertainty in quadrature with the likelihood function
uncertainty leads to a final constraint on the peculiar velocity of
\rxj\ of ${+} 450 \pm 810 \,$km s$^{-1}$.  This constraint is
significantly better than that measured in single clusters measured by,
e.g., SuZIE II \citep{Benson2003}, showing the promise of Z-Spec for
this kind of measurement.

Figure \ref{fig:continuum}\ shows the kSZ amplitude measured here
superimposed on the sub-mm background simulation results discussed in
Section \ref{sS:systematics}.  We note that given the $\sim 1 \,$mJy
amplitude of the kSZ signal inferred in this cluster, the sub-mm
background produces a $\sim 5 \,$\% systematic uncertainty in the
measurement.  This is possibly smaller than might be expected, but we
note that in this case the chopping strategy employed by Z-Spec works
to our advantage, causing the faint, confused sub-mm background to
largely cancel.  Put another way, the faint CIB imposes a limit on the
accuracy to which we can measure the zero point of the measurement
using the control field data, which we estimate as $\sim 50 \, \mu$Jy
beam$^{-1}$.

\section{Discussion}
\label{S:discussion}

The results presented in this work represent the first high spectral
resolution measurements of the SZ effect near its null at $217 \,$GHz.
The SZ null position is measured as $\nu_{0} = 225.8 \pm 2.5
\mathrm{(stat.)} \pm 1.2 \mathrm{(sys.)} \,$GHz, which differs from
the canonical null frequency by $3.0 \sigma$.  By simulating the
response of the instrument to the sky, we measure the best fitting SZ
model to be $y_{0} = 4.6^{+0.6}_{-0.9} \times 10^{-4},
T_{\mathrm{e},0} = 15.2^{+12}_{-7.4} \,$keV for $v_{\mathrm{pec}} = 0
\,$km s$^{-1}$.  When $v_{\mathrm{pec}}$ is allowed to vary, a most
probable value of $v_{\mathrm{pec}} = {+}450 \pm 810 \,$km s$^{-1}$ is
found.  Due to differences in modeling, calibration, and assumptions
regarding the necessity of spectral corrections to the SZ effect, it
is difficult to directly compare the results derived here to those in
the literature.  Our use of the Bolocam data mitigates this within our
analysis since we have full control over the data which image a large
area around the cluster to very low noise levels.  As a cautionary
note, simply varying the $\beta$ model to values found in the
literature and fitting to the Bolocam data used here can lead to
changes of $25$\% in the inferred $y_{0}$ at $140 \,$GHz.  The same
magnitude of variation in $y_{0}$ from the Z-Spec data assuming
different models should be expected.  We therefore hesitate to perform
an in-depth comparison to previous work in \rxj, over which the
$y_{0}$ measured can differ from ours by as much as $\sim 250 \,$\%
(e.g., \citealt{Benson2004}).

Since the Z-Spec constraints are not limited by systematic errors, it
should be possible to reduce the uncertainties of Z-Spec measurements
substantially.  A simple change would be to increase the integration
time, which would reduce the uncertainty as $1/\sqrt{t}$ until the
close to the systematic error floor.  Furthermore, several
inefficiencies arise in these measurements because of the chop and nod
cycle used in the observations.  First, it would be better to chop and
nod at a higher frequency to reduce non-cancellation in the chops from
changing atmosphere, particularly on poor weather days where the data
cut fraction is large.  In addition, a chop waveform closer to a sine
wave, or alternatively choosing a demodulation waveform closer to the
true chop pattern, would increase the in-band signal.  Third,
selecting a larger chop throw would decrease the SZ-signal
attenuation, thereby increasing the signal strength per integration
time.  Finally, chopping in constant right ascension rather than
constant azimuth would allow careful placement of the chopped beams on
faint emission positions.  A methodological improvement unrelated to
the chop would be to acquire more blank sky data to enable more
systematics checks.  Together, instituting these changes would
dramatically increase Z-Spec's sensitivity to diffuse SZ emission and
reduce the uncertainties on both the rSZ and kSZ measurements.  We
expect instituting these changes and surveying several clusters with
both Z-Spec and Bolocam could lead to $v_{\mathrm{pec}}$ constrains
several times better than those found with SuZIE II in single
clusters.

While the first measurement of its kind, the Z-Spec results presented
here yield much more powerful constraints on the SZ spectral
corrections when the Bolocam data are included.  The cause of this is
simple; over a small spectral range like the one afforded by the $220
\,$GHz atmospheric window, the changes in the spectral curvature of
the tSZ from the SZ corrections are always small.  Furthermore, since
rSZ, kSZ, and foreground contamination can all lead to a shift in the
null of the SZ effect, it is impossible to unambiguously disentangle
their contributions from the SZ null shift alone.  For this reason, if
the goal were to measure spectral corrections to the SZ effect, it
would be preferable to observe at many wavelengths.  However,
combining measurements from different instruments is difficult due to
the various cross-calibration uncertainties which exist: it is better
to use a single instrument taking data in several bands at once whose
cross-band calibrations are well understood.  As a general rule, an $R
\sim 5{-}10$ imaging instrument with at least several bands between 15
and $600 \,$GHz would be the best choice, as this optimizes the
trade-off between bandwidth and spectral resolution over the entire SZ
spectrum.  The successful SuZIE series of instruments
\citep{Holzapfel1997a} are an early example of this approach, as are
the MAD \citep{dePetris2007}, OLIMPO \citep{Masi2008}, and the Planck
Surveyor \citep{Planck2011}.  In the near future MUSIC, scheduled to
be deployed at the CSO in 2012, will have four bands with $5 \lesssim
R \lesssim 15$ between 150 and 350~GHz filling a 14~arcmin FOV, which
will enable imaging of the SZ spectrum over entire clusters
\citep{maloney10}.  Further in the future, the next generation of
ground-based mm/sub-mm telescopes placed at high, dry sites like
CCAT\footnote{{\url http://ccatobservatory.org/}} will have cameras
covering two full octaves from $3 \,$mm to $750 \, \mu$m over perhaps
six bands with $R \sim 10$.  An instrument like this would allow a
major step forward in SZ effect spectral studies.

\section*{Acknowledgments} 

Our thanks to M.~Hollister for his help acquiring the Bolocam data
used in this work, J.~Filippini for many useful discussions on
statistics \&c., and an anonymous referee for useful suggestions which
improved this manuscript.  The Z-Spec team acknowledges support from
the following grants for building and fielding the instrument: NASA
SARA grants NAGS-11911 and NAGS-12788, and NSF AST grant 0807990.
Bolocam was constructed and commissioned using funds from
NSF/AST-9618798, NSF/AST-0098737, NSF/AST-9980846, NSF/AST-0229008,
and NSF/AST-0206158.  The Bolocam observation and data analysis
efforts were also supported by the Gordon and Betty Moore Foundation.
J.S.~was partially supported by a NASA Post-doctoral Program
fellowship, NSF/AST-0838261, and NASA/NNX11AB07G; N.C.~was partially
supported by NASA Graduate Student Research Fellowship.  This research
has made use of data obtained from the Chandra Data Archive and
software provided by the Chandra X-ray Center (CXC) in the application
package CIAO.

\bibliography{ms}

\end{document}